\documentclass[english,prd,onecolumn,superscriptaddress,nofootinbib,preprintnumbers]{revtex4}
\usepackage[latin1]{inputenc}
\usepackage{graphicx}
\usepackage{amssymb}
\usepackage{color}
\usepackage{float}
\usepackage{amsmath}
\usepackage{amsfonts}
\usepackage{dcolumn}
\usepackage{hyperref}

\begin{document}

\title{Late time cosmic acceleration: ABCD of dark energy  and modified theories of
gravity\footnote{Dedicated to 75th birthday of J.V. Narlikar.}}

\author{M.~Sami}
 \affiliation{Centre for Theoretical Physics, Jamia Millia Islamia,
New Delhi-110025, India}
\affiliation{Eurasian  International Center
for Theoretical Physics, Eurasian National University, Astana
010008, Kazakhstan}
\author{R.~Myrzakulov}
\affiliation{Eurasian  International Center for Theoretical Physics,
Eurasian National University, Astana 010008, Kazakhstan}

\begin{abstract}
We briefly review the problems and prospects of the standard lore of
dark energy. We have shown that scalar fields, in principle, can not
address the cosmological constant problem. Indeed, a fundamental
scalar field is faced with a similar problem dubbed {\it
naturalness}. In order to keep the discussion pedagogical aimed at a
wider audience, we have avoided technical complications in several
places and resorted to heuristic arguments based on physical
perceptions. We presented underlying ideas of modified theories
based upon chameleon mechanism and Vainshtein screening. We have
given a lucid illustration of recently investigated ghost free non
linear massive gravity. Again we have sacrificed rigor and confined
to the basic ideas that led to the formulation of the theory. The
review ends with a brief discussion on the difficulties of the
theory applied to cosmology.

\end{abstract}

\maketitle

\section{Introduction}
The standard model {\it a la} hot big bang has several remarkable
successes to its credit which include the predictions of  expansion
of universe \cite{Hubble:1929ig}, existence of microwave background
radiation \cite{Penzias:1965wn} and synthesis of light elements in
the early universe \cite{BBN}. There is a definite mechanism for
structure formation in the standard model: tiny perturbations of
primordial nature may grow via gravitational instability into the
structure we see today in the universe. These inhomogeneities were
observed by COBE in 1992 \cite{Smoot:1992td}. The hot big bang model
requires the tiny perturbations for observational consistency and
structure formation but nevertheless lacks a generic mechanism for
their generation. The latter is seen as one of the fundamental
difficulties associated with the standard model. The other
shortcomings include flatness problem, horizon problem and few more
which belongs to the list of logical inconsistencies of the standard
model where as the problem of primordial perturbations is directly
related to observation. The said difficulties are beautifully
addressed by the inflationary paradigm. Interestingly, cosmological
inflation was invented to tackle the logical inconsistencies of hot
big bang. As for density perturbations, it turned out later that
they could be generated quantum mechanically during inflation and
then amplified to the required level  which certainly came as a big
bonus for inflation. It, therefore, became clear around 1982 that
standard model needs to be complimented by an early phase of
accelerated expansion $-$ the
inflation \cite{Guth:1980zm,Starobinsky:1982ee,linde,stein}.\\
There is one more inconsistency of observational nature, the
standard model of universe is plagued with $-$  the age of universe
in the model falls shorter than the age of  some well known objects
in the universe \cite{Krauss:1995yb,Turner:1997de,Krauss}. The age
crisis is related to the late time expansion as universe spent most
of its time in the matter dominated era for the simple reason that
the expansion rate changed fast in the radiation dominated phase. At
early epochs, universe expands fast and particles move away from
each other with enormous velocities; the role of gravity is to
decelerate this motion. Higher is the matter density present in the
universe, less time the universe would spend to reach a given
expansion rate, in particular, the present Hubble rate, thereby
leading to less age of universe. But whatever percentage of matter
we have in the Universe today is an objective reality and we can do
nothing with it. The only known way out in the standard model is
then to introduce a repulsive effect to encounter the influence of
normal matter which could then allow us to improve upon the age of
universe. Thus, we again need an accelerated phase of expansion at
late times to address the age crisis \cite{Krauss:1995yb}. It is
remarkable that the late time cosmic acceleration was directly
observed in 1998 in supernovae Ia observations \cite{SN} and was
confirmed by
indirect observations thereafter \cite{cmb,wmap,Ade:2013zuv}. \\
It is interesting that accelerated expansion plays an important role
in the dynamical history of our universe: the hot big bang model is
sandwiched between two phases of fast expansion $-$ inflation
\cite{Guth:1980zm,Starobinsky:1982ee,linde,stein} and late time
cosmic acceleration \cite{review1,paddy,
vpaddy,review2,review3,review3C,review3d,review4,Clifton:2011jh,smyr,sef}
needed to solve the generic inconsistencies of the standard model of
universe. Late time cosmic acceleration is an observed phenomenon at
present \cite{SN} where as similar
confirmation for inflation is still awaited.\\
In cosmology, observations supersede theoretical model building at
present. What causes late time cosmic acceleration is the the puzzle
of the millennium. There are many ways  of obtaining late time
acceleration \cite{review1,paddy,
vpaddy,review2,review3,review3C,review3d,review4,Clifton:2011jh,smyr,sef}
but observations at present are not yet in position to distinguish
between them. Broadly, the models aiming to address the problem come
in two categories $-$ the standard lore based upon Einstein theory
of general relativity (GR) with a supplement of energy momentum
tensor by an exotic component dubbed {\it dark energy}
\cite{review2} and
scenarios based upon {\it large scale modification of gravity} \cite{Clifton:2011jh}.\\
Which of the two classes of models has more aesthetics is a matter
of taste. Let us first briefly discuss the dark energy scenario. The
simplest model of dark energy is based upon cosmological constant
$\Lambda$ which is an integral part of Einstein's gravity. All the
observations at present are consistent with the model based upon
cosmological constant $-$ $\Lambda CDM$. However, there are
difficult theoretical problems associated with $\Lambda$. With a
hope  to alleviate these problems, one  tacitly switches off
$\Lambda$ without justification and introduces scalar fields with
generic cosmological dynamics which would mimic cosmological
constant at present. Unfortunately, scalar field models are faced
with  problems similar to cosmological constant. As for the standard
lore, to be fair, cosmological constant performs satisfactorily on
observational grounds and unlike scalar fields does not require
 adhoc assumption for its introduction. \\
What goes in favor of modified theories of gravity? Well, Einstein
theory of gravity is directly confronted with observations at the
level of solar system; it describes local physics with great
accuracy and
 is extrapolated with great confidence to large scales where it has
 never been verified directly. We know that gravity is modified at small
distances via quantum corrections, it might be that it also suffers
modification at large scales. And it is quite natural and intriguing
to imagine that these modifications give rise to late time cosmic
acceleration. What kind of modifications to gravity can be expected
at low energies or at large scales? Weinberg theorem tells us that
Einstein gravity is the unique low energy field theory of (massless)
spin 2 particles obeying Lorentz invariance. It is therefore not
surprising that most of the modified theories of gravity are
represented by Einstein gravity plus extra degrees of freedom. For
instance, $f(R)$ \cite{no,Sotiriou:2008rp,DeFelice:2010aj} contains
a scalar degree of freedom with a canonical scalar field uniquely
constructed from Ricci scalar and the derivative of  $f(R)$ with
respect to $R$. A variety of modified schemes of gravity can be
represented  by scalar tensor theories. In this set up, the extra
degrees of freedom normally mixed with the curvature; action can be
diagonalized by performing a conformal transformation to Einstein
frame where they get directly coupled to matter. All the problems of
modified theories stem from the following requirement. The extra
degrees of freedom should to give rise to late time cosmic
acceleration at large scales and become invisible locally where
Einstein gravity is in excellent agreement with observations. Local
gravity constraints pose real challenge to large scale modification
of gravity; spacial mechanisms are required to hide these degrees of
freedom. Broadly there are two ways of suppressing  them locally.
(1) Chameleon screening \cite{Khoury:2003rn,Khoury:2003aq,brax}:
this mechanism is suitable to massive degrees of freedom such that
the masses become very heavy in high density regime allowing to
escape their detection locally. (2) Vainshtein screening
\cite{Vainshtein:1972sx,Babichev:2013usa,Babichev:2009us} suitable
to massless degrees of freedom, operates via kinetic suppression
such that around a massive body, in a large radius known as
Vainshtein radius, thank to non-linear derivative interactions in
the Lagrangian, the extra degrees of freedom gets decoupled from
matter switching off any modification to
gravity locally. \\
In case of massive gravity \cite{deRham:2010ik,deRham:2010kj}, we
end up adding three extra degrees of freedom one of which, namely,
the longitudinal degree of freedom ($\phi$) is coupled to source
with the same strength at par with the zero mode and leads to vDVZ
discontinuity \cite{vanDam:1970vg,Zakharov:1970cc} in linear theory.
In $dRGT$ \cite{deRham:2010ik,deRham:2010kj}, in decoupling limit,
valid limit to tackle the local gravity constraints, the
longitudinal mode gets screened by the non-linear derivative terms
of the field $\phi$ dubbed galileon
\cite{Luty:2003vm,Nicolis:2008in}.\\
Models of large scale modifications based upon chameleon mechanism
are faced with tough challenges: These models are generally unstable
under quantum corrections as the mass of the field should be large
in high density regime in order to pass the local physics
constraints \cite{Upadhye:2012vh,Gannouji:2012iy}. In attempt to
comply with the local physics, one also kills the scope of these
theories for late time cosmic acceleration\cite{scope}. On the other
hand, Vainshtein mechanism is a superior field theoretic method of
hiding extra degrees of freedom and is at the heart of recently
formulated ghost free model of massive gravity$-$ $dRGT$. Apart from
the superluminality problem \cite{Deser:2012qx} of $dRGT$ inherent
to galileons \cite{Nicolis:2008in,Hinterbichler:2009kq,Goon:2010xh}
, it is quite discouraging that there no scope of
Freidmann-Robertson-Walker (FRW) cosmology in this theory
\cite{D'Amico:2011jj}. It is really a challenging task to build a
consistent theory of massive
gravity with a healthy cosmology.\\
In this paper, we shall briefly review the problems associated with
dark energy and focus on problems and prospects of modified theories
of gravity and their relevance to late time cosmic acceleration. The
review is neither technical nor popular, it is rather a first
introduction to the subject and aims at a wider audience.\\
In this review, we would stick to metric signature,$(-,+,+,+)$ and
denote the reduced Planck mass as $M_p=(8 \pi G)^{-1/2}$. We hereby
give an unsolicited advise to the reader on the follow up of the
review. The section on cosmological constant should be complemented
by the Ref.\cite{ahmad} for a thorough understanding of the problem.
For a detailed study of scalar field dynamics, we refer the reader
to the review\cite{review2}. Reader interested in learning more on
modified theories of gravity,  supported by chameleon mechanism, is
recommended to work through the
reviews\cite{DeFelice:2010aj,brax,khoury}. In our description of
massive gravity, we resorted to heuristic arguments in several
places in order to avoid the technical complications. After reading
the relevant section, we refer  the reader to the exhaustive
reviews\cite{Hinterbichler:2011tt,claudia} on the related theme.

\section{FRW cosmology in brief}
The Friedmann-Robertson-Walker  model is based on the assumption of
homogeneity and isotropy {\it a la} cosmological principal
\footnote{The  standard or restricted cosmological principle deals
with homogeneity and isotropy of three space. The success of hot big
bang based upon this doctrine  witnesses that not always nature
makes choice for the most beautiful. On the other hand, the perfect
cosmological principal, in adherence to the fundamental principle of
relativity, treats space and time on the same footings. It imbibes
aesthetics, beauty and is certainly on a solid philosophical ground
than the restricted cosmological principle. Interestingly, the
nineteenth century  materialist philosophy$-$ the dialectical
materialism view on the genesis of universe was based upon a similar
principal which can be found in the classic work  by Frederick
Engels, " Dialectics of nature". According to this ideology ,
universe is infinite, had no beginning, no end and always appears
same thereby leaving no place for God in it. The Hoyle-Narlikar
steady state theory  is based upon the perfect cosmological
principle and it would have been extremely pleasing had the  study
state theory succeeded but we can not force nature to make a
particular choice, even the most beautiful one! }which is
approximately true at large scales. The small deviation from
homogeneity in the early universe seems to have played very
important role in the dynamical history of our universe. The tiny
density fluctuations are believed to have grown via gravitational
instability into the
structure we see today in the universe. \\
Homogeneity and isotropy forces the metric of space time to assume
the form,
\begin{equation}
\label{frw}
ds^2=-dt^2+a^2(t)\left(\frac{dr^2}{1-Kr^2}+r^2(d\theta^2+\sin^2\theta
d\phi^2)\right);~~K=0,\pm 1
\end{equation}
where $a(t)$ is scale factor. Eq(\ref{frw}) is purely a kinematic
statement which is an expression of maximal spatial symmetry of
universe thanks to which full information of cosmological dynamics
is imbibed in a single function$-$ a(t). Einstein equations allow us
to determine the scale factor provided the matter contents of
universe are specified. Constant $K$ occurring in the metric
(\ref{frw}) describes the geometry of spatial section of space-time.
Its value is also determined once the matter distribution in the
universe is known.  In general, Einstein equations
\begin{equation}
R^\mu_\nu-\frac{1}{2}\delta^\mu_\nu R=8\pi GT^\mu_\nu
\end{equation}
are complicated but thank to the maximal symmetry, expressed by
(\ref{frw}), get simplified and give rise to the following evolution
equations,
\begin{eqnarray}
\label{Freq} && H^2\equiv \frac{\dot{a}^2}{a^2}=\frac{8\pi G
\rho}{3}-\frac{K}{a^2}\\
 &&\frac{\ddot{a}}{a}=-\frac{4\pi G}{3}\left(\rho+3p\right)
\label{aceq}
\end{eqnarray}
where $\rho$ and $p$ are density and pressure of matter filling the
universe which satisfy the continuity equation,
\begin{equation}
\label{cont}
 \dot{\rho}+3H(\rho+p)=0
\end{equation}
For cold dark matter, $p_m=0$ (equation of state parameter
$w_m\equiv p_m/\rho_m=0$) and it follows from (\ref{cont}) that
$\rho_m=\rho_m^0(a_0/a)^3$, where  the subscript "0" designate the
respective quantities at the present epoch. In case of spatially
flat universe, $K=0$, the scale factor $a_0$ can be normalized to
{\it a priori} given value, say at unity. In other cases, its value
depends on the matter
content in the universe.\\

 The nature of expansion expressed by the Equations
(\ref{Freq})$\&$ (\ref{aceq})
 depends upon the nature of the matter content of
universe. It should be emphasized that in general theory of
relativity, pressure contributes to energy density and the latter is
a purely  relativistic effect. The contribution of pressure in
Eq(\ref{aceq}) can qualitatively modify the expansion dynamics.
Indeed, Eq.(\ref{aceq}) tells us that
\begin{eqnarray}
&& \ddot{a}>0,~~~p<-\frac{\rho}{3}\nonumber \\
&& \ddot{a}<0,~~~p>-\frac{\rho}{3}\nonumber
\end{eqnarray}

Accelerated expansion, thus, is fueled by an exotic form of matter
of large negative pressure $-${\it dark energy} \cite{review1,
vpaddy,review2,review3,review3C,review3d,review4,Clifton:2011jh}
which turns gravity into a repulsive force. The simplest example of
a perfect fluid of negative pressure is provided by cosmological
constant associated with $\rho_\Lambda= const$. In this case the
continuity equation (\ref{cont}) yields the relation $p_{\Lambda}
=-\rho_{\Lambda}$. Keeping in mind the late time cosmic evolution,
let us write down the evolution equations in matter dominated era in
presence of cosmological constant,

\begin{eqnarray}
\label{heq1}
 && H^2\equiv \frac{\dot{a}^2}{a^2}=\frac{8\pi G
\rho_m}{3}-\frac{K}{a^2}+\frac{\Lambda}{3}\\
&& \frac{\ddot{a}}{a}=-\frac{4\pi G}{3}\rho+\frac{\Lambda}{3}
\label{aceq1}
\end{eqnarray}

It is instructive to cast these equations in the form to mimic the
motion of a point particle in one dimension. Eq.(\ref{aceq1}) can be
put in the following form,
\begin{equation}
\ddot{a}(t)=-\frac{\partial V(a)}{\partial
a};~V(a)=-\left(\frac{4\pi G\rho_b a^2}{3}+\frac{\Lambda
a^2}{6}\right)
\end{equation}
whereas the Friedmann equation acquires the form of total energy of
the mechanical particle,
\begin{equation}
E=\frac{\dot{a}^2}{2}+V(a) ,~ E=-\frac{K}{2}
\end{equation}
\begin{figure}[h]
\includegraphics[scale=0.4]{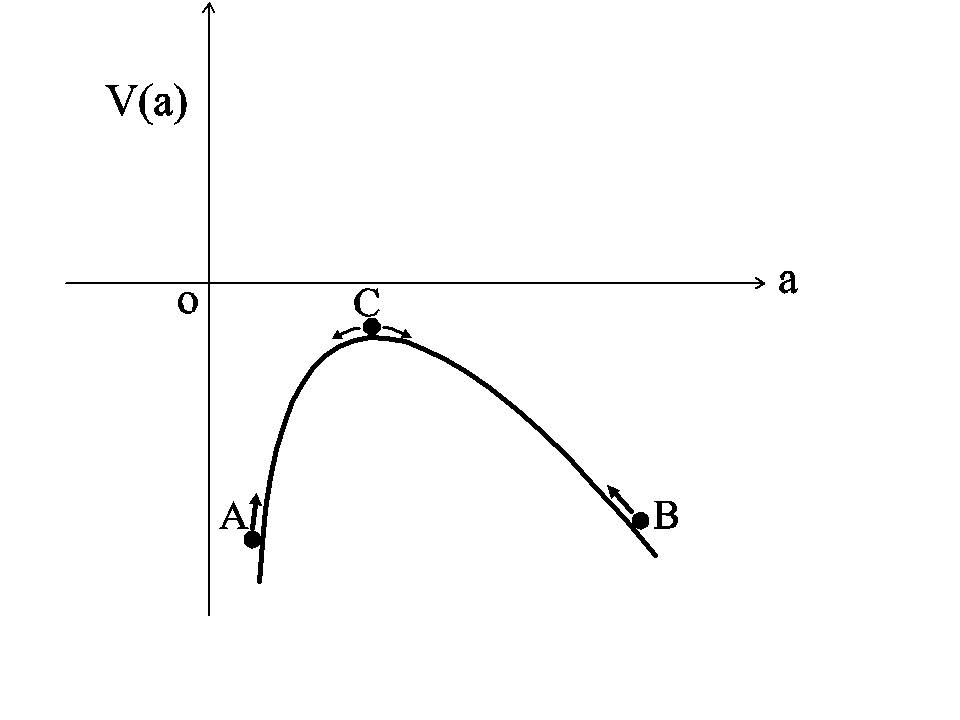}
\caption{ Figure displays the potential V (a) versus the scale
factor $a$ . The initial positions (A) and (B) correspond to motion
of system beginning from a = 0 and $a =\infty$. For $K<0$ and
$\Lambda<\Lambda_c$, we have oscillating and bouncing solutions
depending whether the motion commences from configuration (A) with
$a=0$ or from configuration (B) with $a=\infty$.
 Einstein static solution($\ddot{a}=0,\dot{a}=0$) corresponds
to the maximum of the potential. The case of $\Lambda>\Lambda_c$ is
similar to $K=0,-1$ such that the kinetic energy is always
sufficient to overcome the barrier.}
\label{potc}
\end{figure}
The potential $V(a)$ is concave down and has a maximum where the
kinetic  energy is minimum (see, Fig.\ref{potc}),
\begin{equation}
\left(\frac{\dot{a}^2}{2}\right)_{|_{min}}=\frac{1}{2}\left(C^{2/3}\Lambda^{1/3}-K\right)
\end{equation}
where $C=4\pi G\rho^0a_0^3$. If we imagine that motion in
Fig.\ref{potc} commences on the left of the hump, the kinetic energy
is always sufficient to overcome the barrier for $K=0$ and $K=-1$
where as in case of $K=1$, we get a bound on the value of
$\Lambda\geq \Lambda_c=4\pi\rho^0 a_0^3$ to achieve the same.
Observations have repeatedly conformed the spatially flat nature of
geometry $(K = 0)$ \cite{SN,cmb,wmap} which is consistent with the
prediction of inflationary scenario and we shall adhere to the same
in the following discussion. In this case, starting from position
(A), see Fig\ref{potc}, one can always reach (C) and before one
reaches the hump, motion decelerates followed by acceleration
thereafter. Observations have shown that this transition takes place
at late times. In order to appreciate it, let us  write (\ref{heq1})
in the form,
\begin{equation}
H^2=H^2_0\left[\Omega_m\left(\frac{a_0}{a}\right)^3+\Omega_\Lambda\right];~
\Omega_m=\frac{\rho_m^0}{\rho_{cr}},~\Omega_\Lambda=\frac{\rho_\Lambda}{\rho_{cr}},~\rho_{cr}=\frac{3H_0^2}{8\pi
G}
\end{equation}

It then straight forward to estimate the numerical value of $a_0/a$
for which the kinetic energy,
\begin{equation}
\frac{\dot{a}^2}{2}=H_0^2\left[\Omega_m\left(\frac{a_0}{a}\right)^3+\Omega_\Lambda
a^2\right]
\end{equation}
is minimum and that happens when,
\begin{equation}
\left(\frac{a_0}{a}\right)|_{min}\equiv
1+z_{tr}=\left(\frac{2\Omega_\Lambda}{\Omega_m}\right)^{1/3}
\end{equation}
where we have introduced redshift $z$ which quantifies the effect of
expansion. Using the observed values of dimensionless density
parameters, $\Omega_m\simeq 0.3$ and $\Omega_\Lambda\simeq 0.7$, we
find that $z_{tr}\simeq 0.67$ which tells us that transition from
deceleration to acceleration, indeed, took place
recently.\\

Let us note that cosmological constant is not the only example of
negative pressure fluid, a host of scalar field systems can also
mimic a negative pressure fluid. An important comment about negative
pressure systems is in order. The introduction of $\Lambda$ does not
require an adhoc assumption, the latter is always present in
Einstein equations by virtue of Bianchi identities. In fact in four
dimensions, the only consistent modification(without invoking the
extra degrees of freedom)that the Einstein equations allow in the
classical regime is given by, $T_{\mu\nu}\to T_{\mu\nu}- \Lambda
g_{\mu\nu}$. Actually, this is the other way around that one should
provide justification if one wishes to drop the cosmological
constant from Einstein equations; there exists no symmetry at low
energies to justify the latter. As for the scalar fields, their
introduction is quite adhoc and on the top of every thing, one
switches off $\Lambda$ for no known reason. Scalar fields, however,
may be of interest if they are inspired by a fundamental theory of
high energy physics.
\subsection{Age crisis in hot big bang and the need for a repulsive
effect} At early epochs, radiation dominates, its energy density is
large, as a result, the expansion rate is also large. Consequently,
it does not take much time to reach a given expansion rate in the
early universe. For instance, Universe was around $10^5$ years old
at the radiation matter equality which is negligible  compared to
the age of universe. It is therefore clear that most of the
contribution to the age of universe comes from matter dominated era
at late stages. In order to appreciate the role of $\Lambda$, let us
switch it off in the Friedmann equation. Then for matter dominated
Universe ($K=0$), the Friedmann equation (\ref{Freq})  readily
integrates to,
\begin{equation}
a(t) \propto t^{2/3} \to H=\frac{2}{3t}
\end{equation}
and specializing to the present epoch, we have
\begin{equation}
\label{Age}
 t_0=\frac{2}{3}\frac{1}{H_0}
\end{equation}
 Recent
observations reveal that
\begin{equation}
H^{-1}_0\simeq 1.4\times 10^{9} years\to t_0 \simeq 9.4\times 10^{9}
years
\end{equation}
which falls much shorter than the age of some well known objects
(around 14 billion years) in the universe
\cite{Krauss:1995yb,Turner:1997de,Krauss}. Actually, the factor of
$2/3$ in (\ref{Age}) spoils the estimate. Let us argue on physical
grounds as how to address the problem. In presence of normal matter,
gravity is attractive and it decelerates the motion. If gravity
could be ignored, then using the Hubble law, $v=Hr$($v=const$), we
could have, $t_0=1/H_0$ which is what is required. However, we can
not ignore gravity, there is around 30 $\%$ of matter present in the
universe which causes deceleration of the expansion and reduces the
age of universe. The only way out to decrease the influence of the
matter is to introduce a repulsive effect necessary to encounter the
gravitational attraction of normal matter. Let us stress that this
is the only known possibility to improve upon the age of universe in
the standard model of Universe. Indeed using the Friedmann equation,
we can estimate the time universe has spent starting from the big
bang till today or the age of universe $t_0$,

  \begin{equation}
  \label{tage}
t_0=\frac{1}{H_0}\int_0^\infty{\frac{dz}{(1+z)\sqrt{\Omega_m(1+z)^3+\Omega_{\Lambda}}}}
\end{equation}
 where we have used the change of variable, $ dt=-dz/H(1+z)$. The age of the universe is then finally given by,
\begin{equation}
\label{agel}
t_0H_0=\frac{2}{3}\frac{1}{\Omega_{\Lambda}^{1/2}}\ln\left(
\frac{1+\Omega_{\Lambda}^{1/2}}{\Omega_m^{1/2}}\right)
\end{equation}
 Expression (\ref{agel})  tells us that $t_0H_0\simeq
1$ for the observed values of density parameters,
$\Omega_\Lambda\simeq 0.7$ and $\Omega_m\simeq 0.3$. Thus the {\it
late time inconsistency of hot big bang cries for
cosmological constant}.\\
{ It is really interesting to note that there exists no such problem
in Hoyle-Narlikar steady state cosmology \cite{Hoyle1,Hoyle2} which
thanks to the perfect cosmological principal has no beginning and no
end. Also, the study state theory imbibes cosmic acceleration and
does not suffer from the logical inconsistencies the standard model
is plagued with. Unfortunately, the model faces problems related to
thermalization of the microwave background radiation. However, the
generalized steady state theory dubbed "Quasi Steady State
Cosmology" (QSSC) formulated by Hoyle, Burbidge and Narlikar claims
to explain the CMBR as well as derive its present temperature which
the big bang cannot do\cite{jvnb}.}
\subsection{ Theoretical issues associated with cosmological constant}
It is clear from the aforesaid that cosmological constant is
essentially present in Einstein equations as a free parameter which
should be fixed by observations. Sakharov pointed out in 1968
\cite{Sakharov:1967pk} that quantum fluctuations would correct this
bere value. In flat space time, according to Sakharov, a field
placed in vacuum would have energy momentum tensor
\begin{eqnarray}
<0|T_{\mu \nu}|0>=-\rho_v\eta_{\mu \nu}
\end{eqnarray}
 uniquely fixed by relativistic invariance. $\rho_v$ dubbed vacuum energy density
  is constant by virtue of conservation of energy momentum tensor.
  Keeping in mind the perfect fluid form of the energy momentum tensor, we have, $p_v=-\rho_v$ which is the
  expression of relativistic invariance.
 The curved space time
 generalization is given by
\begin{eqnarray}
<0|T_{\mu\nu}|0>=-\rho_v g_{\mu \nu}
\end{eqnarray} which should be added to the bare value of
cosmological constant present in Einstein equations,
\begin{equation}
\label{eqeq}
 R_{\mu\nu}-\frac{1}{2}R g_{\mu \nu}+g_{\mu
\nu} \Lambda_b= T_{\mu\nu}^m + <0|T_{\mu\nu}|0>
\end{equation}

A free scalar field is an infinite collection of non interacting
harmonic oscillators whose zero point energy is the vacuum energy of
the scalar field,
\begin{equation}
\label{vew}
 \rho_v=\frac{1}{2}\int_0^\infty{\frac{4\pi k^2 dk}{(2\pi)^2}
\sqrt{ k^2+m^2}}
\end{equation}
and incorporating spin does not change the estimate. Expression
(\ref{vew}) is formally divergent and requires a cut off. One
normally cuts it off at Planck's scale as an expression of our
ignorance and concludes that $\rho_v \sim M^4_p$. Using then the
Friedmann equation expressed through dimensionless density
parameters,

\begin{eqnarray}
\Omega^{eff}_{\Lambda}+\Omega_m=1
\end{eqnarray}
one finds, $ \rho_{\Lambda}^{eff}\lesssim \rho_{cr}\sim
10^{-120}M^2_p$ which is the source of a grave problem. And since,
\begin{equation}
\label {effL }
\rho^{eff}_{\Lambda}=\rho^b_{\Lambda}+\rho_v,
\end{equation}
  it follows that
   $\rho^b_\Lambda$ should cancel $\rho_v$
to a fantastic accuracy, typically, at the level of one part in
$10^{-120}$. The supernovae Ia observation in 1998 revealed that
effective vacuum energy is not only small, it is of the order of
matter density today.

The cosmological constant problem is often formulated as,\\
{ \bf $\bullet$}~{\bf Old
problem(before 1998)}: Why effective vacuum energy is so small today?\cite{Wg}, \\
 { $\bullet$}~{\bf New problem(after 1998)}: Why we happen to live
in special times when dark energy density is of the order of matter
density? {\it a
la} coincidence problem \cite{Steinhardt:1999nw}.\\

 We should point out a flaw in
the above arguments\cite{ahmad}. We should bear in mind that the cut
off used on 3-momentum violates Lorentz invariance and might lead to
wrong results. In what follows, we shall explicitly demonstrate it.

Lorentz invariance signifies  a particular relation between vacuum
energy $\rho_v$ and vacuum pressure $p_v$, namely, $\rho_v=-p_v$.
Similar to the vacuum energy, the vacuum pressure is formally
divergent and also requires a cut off. Introducing a cut off $M$ in
the divergent integrals and expressing $\rho_v$ and $p_v$, we have,
\begin{eqnarray}
&&\rho_v=\frac{1}{2(2\pi)^3}\int_0^\infty{d^3{\bf k}~\omega(k)}\\
&&p_v=\frac{1}{6(2\pi)^3}\int_0^{\infty}{d^3{\bf k}\frac{k^2}{w(k)}
};~~w(k)=\sqrt{k^2+m^2}
\end{eqnarray}
 which allows us to compute these quantities,
\begin{eqnarray}
\label{rhov} \label{rhov}
 &&\rho_v=\frac{1}{4\pi^2}\int_0^M
{dkk^2\sqrt{k^2+m^2}}=\frac{M^4}{16
\pi^2}\left[\sqrt{1+\frac{m^2}{M^2}}
\left(1+\frac{m^2}{2M^2}\right)-
\frac{1}{2}\frac{m^4}{M^4}\ln\left(\frac{M}{m}
+\frac{M}{m}\sqrt{1+\frac{m^2}{M^2}}\right)\right]\\
&&p_v=\frac{1}{3}\frac{1}{4\pi^2}\int_0^M{
dk\frac{k^4}{\sqrt{k^2+m^2}}}
=\frac{1}{3}\frac{M^4}{16\pi^2}\left[\sqrt{1+
\frac{m^2}{M^2}}\left(1-\frac{3M^2}{2M^2}\right)+\frac{3m^4}{2M^4}\ln\left(\frac{M}{m}+\frac{M}{m}\sqrt{
1+\frac{m^2}{M^2}}\right)\right] \label{pv}
\end{eqnarray}
In the expressions quoted above, $m$ is the mass of the scalar field
placed in vacuum. Invoking spin contribution does not alter the
estimates. Hence $\rho_v$ and $p_v$ given by (\ref{rhov}) $\&$
(\ref{pv}) are valid estimates for any field placed in vacuum.
Secondly,
 as mentioned before,  for Lorentz invariance to hold we
should have $\rho_v=-p_v$ which is clearly violated by the first
terms in (\ref{rhov}) and (\ref{pv} ). It should be noted that this
is the  first term in these expressions which gives contribution
proportional to $M^4$. As for the second terms with logarithmic
dependence on the cut off, they are in accordance
with Lorentz invariance. \\
It is therefore clear that we should employ a regularization scheme
which respects Lorentz invariance. For instance, dimensional
regularization is suitable to the problem. Let us first transform
the integral from four to $d$ dimension,
\begin{eqnarray}
&&\rho_v=\frac{\mu^{4-d}}{2(2\pi)^{(d-1)}}\int_0^\infty
{dkk^{d-2}d^{d-2}\Omega~\omega(k)}
\end{eqnarray}
where the scale $\mu$ is introduced to take care of the units in $d$
dimensional case and $\Omega$ is the solid angle . This integral can
expressed through gamma function,
\begin{equation}
\rho_v=\frac{\mu^4}{2(4\pi)^{(d-1)/2}}\frac{\Gamma(
-d/2)}{\Gamma(-1/2)}\left(\frac{m}{\mu}\right)^d
\end{equation}
Finally, we should return to four dimensions by letting
$d=4-\epsilon$ and expanding the result in $\epsilon$ to the leading
order,

\begin{equation}
\rho_v =
-\frac{m^4}{64\pi^2}\left(\frac{2}{\epsilon}+\frac{3}{2}-\gamma
-\ln\left(\frac{m^2}{4 \pi \mu^2}\right)\right)+..
\end{equation}
which diverges as $\epsilon\to 0$. We have successfully isolated the
divergence without violating Lorentz invariance. We then subtract
out infinity to obtain the final result,
\begin{eqnarray}
\rho_v\simeq \frac{m^4}{64\pi^2}\ln
 \left(\frac{m^2}{\mu^2}\right)
 \end{eqnarray}
In order to estimate the vacuum energy, we should imagine all the
fields placed in vacuum and sum up their contributions.
  To be
 pragmatic, we use the following data from standard model of particle
 physics to estimate $\rho_v$
 \begin{eqnarray}
m_t\simeq 171 GeV; m_H \simeq 125 GeV; ~~m_{z,w}\simeq 90 GeV,...
\end{eqnarray}
Clearly the stage is set by the heaviest scale in the problem, the
mass of the top quark. As for the scale $\mu$, it is always
estimated by the physical conditions. In the problem under
consideration, the energy scale, $\mu$, is set by the critical
energy density and the energy density characterized by the
wavelength of light received from supernovae,
\begin{eqnarray}
 && \mu \sim \sqrt{H_0 E_\gamma} , H_0 \sim 10^{-41}
GeV;~\lambda\sim 500nm\\
&&\mu \sim \sqrt{H_0 E_\gamma}\to  \rho_v \simeq 10^{8} GeV^4
\end{eqnarray}
which shows that effective vacuum energy density is down by sixty
four orders of magnitude compared to the one obtained using the
Lorentz violating regularization. And this considerably reduces the
fine tuning at the level of standard model,
\begin{equation}
\rho^{eff}_\Lambda\simeq 10^{-56} M_p^4
\end{equation}

 Thus fine tuning is one
part in $10^{-56}$, rather than  one part in $10^{-120}$ as often
quoted, provided we believe that there is no physics beyond standard
model. But we know that there is at least one scale beyond,
associated with gravity, namely, the Planck scale which would take
us back to original fine tuning problem if the Planck scale is
fundamental. However, if it is a derived scale similar to the one in
Randall-Sundrum scenario, the fine tuning could considerable reduce.
We thus conclude that the cosmological constant problem {\it a la}
fine tuning could not be as severe as it
is posed; it is often over emphasized. Of course the problem still remains to be grave.\\
The coincidence problem or why dark energy density is of the order
of matter density today? is yet more over emphasized. We know that
universe went through a crucial transition between $z=1$ to $z=0$.
Let us ask how much time universe has spent beginning from a given
redshift $z$ to the present epoch. Using Eq.(\ref{tage}), it is
straight forward to write down the expression for $t_z$,
\begin{equation}
t_z=\frac{1}{H_0}\int_0^z{\frac{dz'}{(1+z')\sqrt{\Omega_m(1+z')^3+\Omega_{\Lambda}}}}
\end{equation}
where the dimensionless density parameters are specialized to the
present epoch as before.
\begin{figure}[h]
\includegraphics[scale=0.4]{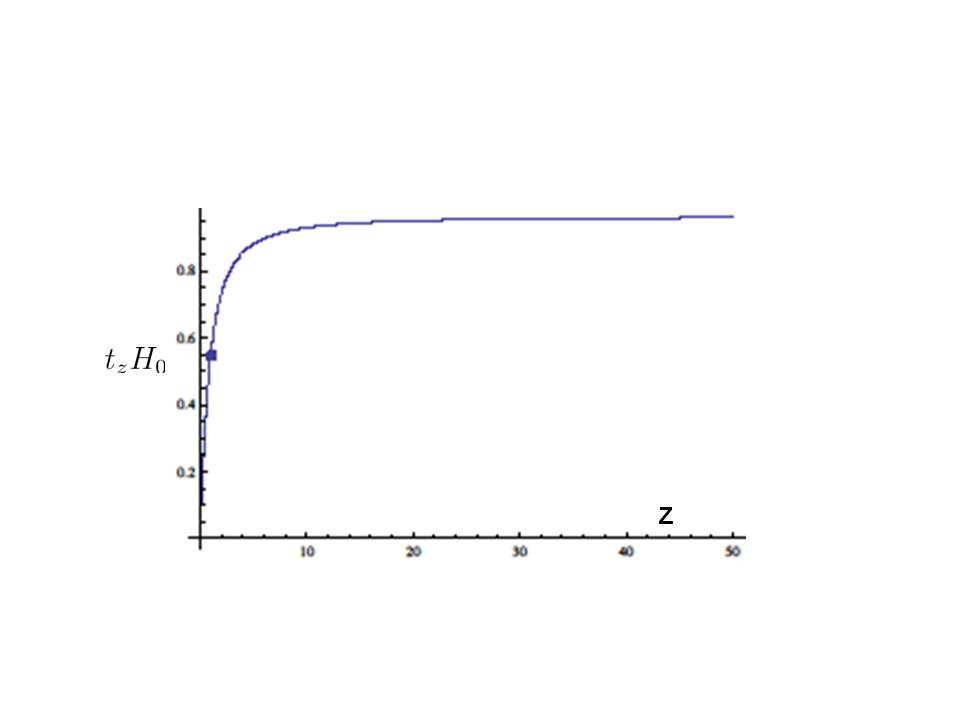}
\caption{Figure shows the time universe  has spent beginning from
particular value of redshift to the present epoch. The black dot on
the curve corresponds to time universe has spent from $z=1$ to $z=0$
which is more than half of the age of universe. It is clear from the
figure that the curve fast saturates as $z$ increases and that most
of the contribution to the age comes from the matter dominated era,
(z=4,0)} \label{age}
\end{figure}

It is clear from the Fig.\ref{age} that most of the contribution to
age comes from late stage of evolution. Universe spent more than
half of its age in the interval between $z=1$ and the present epoch,
$z=0$ and during this period matter density and dark energy density
 remained roughly within the same order of magnitude. Thus they
have been within one order of magnitude for ages thereby telling us
that there is hardly  any {\it   coincidence problem}\cite{carlo}.

\section{Quintessence and its difficulties}
Slowly rolling scalar fields, broadly referred to as ${\it
quintessence}$ \cite{quint}, were introduced with a hope to
alleviate the fine tuning problem. Scalar field models applied to
cosmological dynamics can be classified into two types $-$ {\it
trackers} \cite{Steinhardt:1999nw} and {\it thawing}
\cite{Caldwell:2005tm} models. Trackers are interesting for the
reason that dynamics in this case is independent of initial
conditions where as the thawing models involve dependency on initial
conditions with the same level of fine
tuning at par with cosmological constant. \\

Let us briefly consider the cosmological dynamics of a scalar field
which can be treated as a perfect fluid with energy density
$\rho_\phi$ and pressure $p_\phi$ given by(see Ref.\cite{review2}
for details),
\begin{equation}
\rho_\phi=\frac{\dot{\phi}^2}{2}+V(\phi)~~;
p_\phi=\frac{\dot{\phi}^2}{2}-V(\phi);~~\omega_\phi\equiv
\frac{p_\phi}{\rho_\phi}
\end{equation}
For slowly evolving field, $\omega_\phi \simeq -1$ whereas
$\omega_\phi \simeq 1$ if field rolls  fast which happens for a
steep potential.
 The equation of motion for
the standard scalar field $\phi$ in FRW cosmology is,
\begin{equation}
\label{feq} \ddot{\phi}+3H\dot{\phi}+V'(\phi)=0
\end{equation}
where the second term is due to Hubble damping. From (\ref{feq}), we
infer that
\begin{equation}
\rho_{\phi}=\rho_{\phi}^0
\exp{\left(-\int{3(1+\omega_{\phi})\frac{da}{a}}\right)}
\end{equation}
which tells us that $\rho_\phi \sim 1/a^6$ in case the field is
rolling along a steep potential. Let us consider an exponential
potential which has served as a laboratory for the understanding of
cosmological dynamics \cite{Copeland:1997et,Nunes:2000yc},
\begin{equation}
V(\phi)=V_0 e^{-\lambda \phi/M_p}
\end{equation}

\begin{figure}[h]
\includegraphics[scale=0.4]{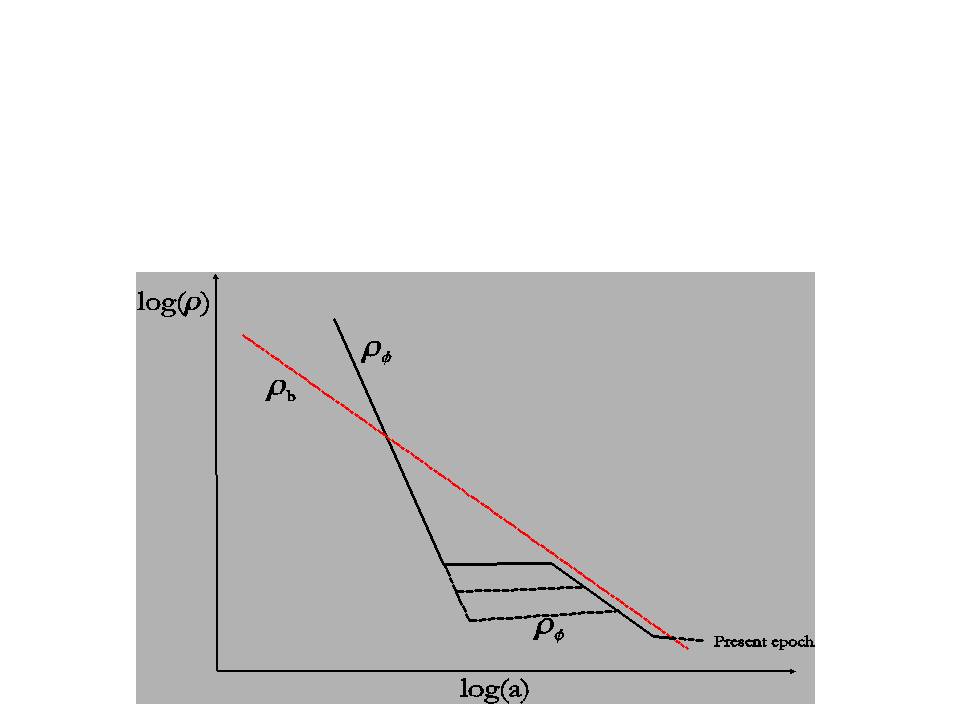}
\caption{Figure shows evolution of $\rho_{\phi}$ and background
energy density versus the scale factor on the logarithmic scale. As
the field emerges from locking regime, it tracks the background. At
late times, it begins to approach $\rho_b$ and finally overtakes to
become dominant giving rise to late time acceleration. Once the
present epoch is set by suitably choosing the model parameters,
evolution is independent of initial conditions.} \label{tracker}
\end{figure}
\begin{figure}[h]
\includegraphics[scale=0.4]{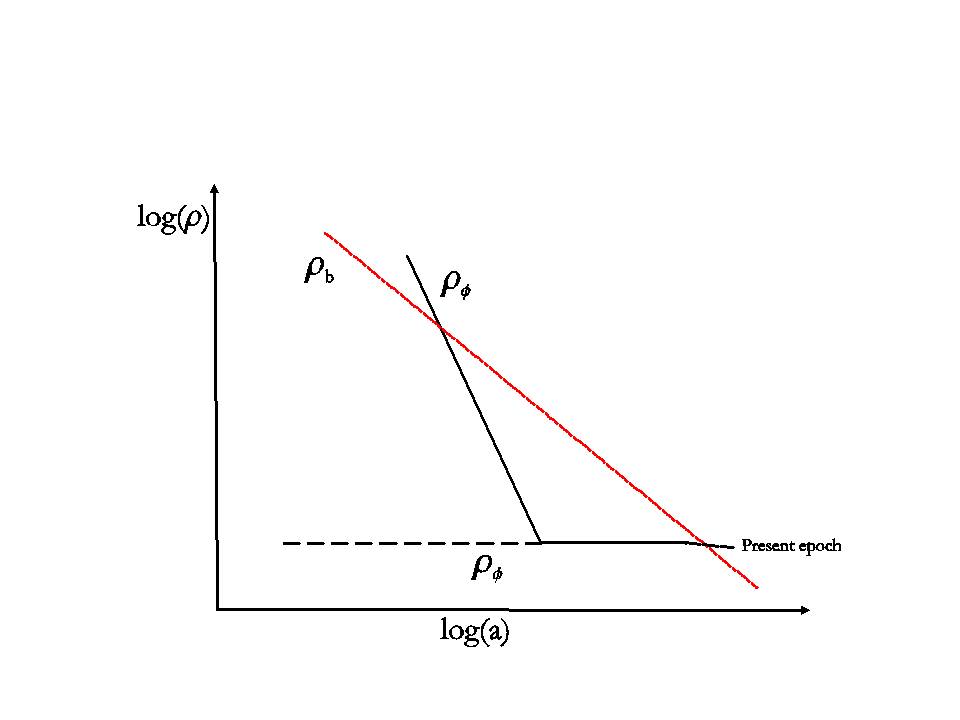}
\label{ntracker} \caption{ The figure shows the evolution of field
energy density along with the background matter density $\rho_b$ on
log scale. Initially, field rolls along steep part of the potential,
redshifts faster than $\rho_b$, overshoots it and freezes due to
large Hubble damping. In this case, after the exit from locking
regime, field begins to roll slowly and overtakes the background and
can account for late time cosmic acceleration. In this case changing
initial conditions would disturb present day physics which can be
restored by resetting the model parameters. In this case, evolution
depends upon initial conditions. The level of fine tuning is at par
with cosmological constant} \label{ntracker1}
\end{figure}

The  parameter $\epsilon=M^2_p (V'/V)^2/2$ then sets a condition for
slow roll, namely, $\lambda<\sqrt{2}$. The slow roll parameters do
not play the same role  here as they do in case of inflation due to
the presence of matter  but still can guide us for the broad
picture.
 A suitable choice of  $\lambda$ can give rise to viable late time cosmic
evolution.  The de Sitter solution is an attractor of the system.
There is one more remarkable attractor in the system that exists in
presence of background (matter/radiation) dubbed scaling solution
which exists for a steep potential with $\lambda \geq \sqrt{3}$. Let
us consider the case when field energy density is initially larger
than the background energy density, $\rho_b=\rho_r/\rho_m$, see
Fig.\ref{tracker}. Since the potential is steep, $\rho_\phi$
redshifts faster than $\rho_b$ and the field overshoots the
background such that $\rho_\phi<<\rho_b$. In that case, the Hubble
damping in the field evolution equation is enormous and
consequently, the field freezes on its potential such that
$\rho_\phi=const$. Meanwhile the background energy density redshifts
with the expansion and the field waits till the moment its energy
density becomes comparable to that of the background, thereafter the
evolution can proceed in two ways depending upon the nature of the
potential: (1) : In case of (steep) exponential potential, field
would track the background; in matter dominated era, field would
mimic matter $(\omega_\phi=\omega_m)$ for ever. This is a very
useful attractor dubbed {\it scaling solution} though not suitable
to late time acceleration. In this case, we shall need a feature in
the potential that would give rise to the exit from scaling solution
at late times, see Fig.\ref{tracker}. (2) In this case, field begins
to evolve and overtakes the background without following it which
happens if the field rolls slow at late times. This happens in case
of
 a potential
which is steep but not exponential at early epochs and shallow at
late times. For such potentials, evolution crucially depends upon
the initial conditions. In this case, though we can have suitable
late time evolution but the model is faced with the same fine tuning
problem as the one based upon cosmological constant; models with
shallow potential throughout are faced with the same problem. Models
of this class are termed as thawing models, see Fig.\ref{ntracker1}.
Let us note that the requirement to obtain a tracker solution is
very specific and only a small number of field potentials in case of
a standard scalar field can give rise to tracker solutions. As for
the tachyon\cite{samio} or
phantom\cite{phantom1,phantom2}\footnote{Phantom field is nothing
but Hoyle-Narlikar creation field $C$   needed in study state theory
to reconcile with homogeneous density by creation of new matter in
the voids caused by the expansion of the universe thereby allowing
the Universe to appear same all the times. } fields, there exists no
realistic tracker (that could tracker the standard matter);
irrespective of their potential they belong to the class of thawing
models.\\

What is a desirable quintessence field for thermal history and late
time cosmic evolution? Actually, we should look for a model with
steep exponential potential throughout most of the history of
universe and a shallow one  at late times. In that case, the field
would assume the scaling behavior after the exit from locking regime
and only at late times it would leave it to become dominant and give
rise to late time cosmic evolution {\it a la tracker solution}, see
Fig.\ref{tracker} \cite{Steinhardt:1999nw}. In this case, evolution
is independent of initial conditions and the fine tuning associated
with $\Lambda$ may be alleviated. It is possible to realize tracker
solutions in several ways. However, they are obtained most naturally
in models with inverse power law potentials ($V\sim 1/\phi^n$) which
approximate the exponential potential for large values of the
exponent $n$ and for which the slope is variable$-$ large at early
epochs and small at late times which is precisely the behavior we
are looking for.
 It is
little discouraging that tracker models are less favored
observationally compared to
thawing models.\\

The slowly rolling scalar field models irrespective of their types
are generally faced with another grave problem which surfaces when
we allow the scalar field interaction with matter, $g \phi
\bar{\psi}\psi$. In order to appreciate the problem, let us estimate
the mass of scalar field employing any of the slow roll conditions,
\begin{eqnarray}
&& \epsilon=\frac{M^2_p}{2}\left(\frac{V'}{V}\right)^2 <<1 ;~ \eta=
M^2_p \frac{V''}{V}<<1
\end{eqnarray}
and since the mass of the field should be of the order of $H_0$ to
be relevant to late time cosmic acceleration, we find by making use
of the second slow roll parameter $\eta$,
\begin{eqnarray}
m^2 \simeq V'' \simeq \frac {V}{M^2_p} \simeq \frac {H^2_0
M^2_p}{M^2_p} \simeq\left( 10^{-33}eV\right)^2
\end{eqnarray}
An important remark related to late time field dynamics is in order.
In case, $m>>H_0$, the field would be rolling very fast at the
present epoch and hence of no relevance to late time cosmology. On
the other hand, if $m<<H_0$, the field would not be distinguished
from cosmological constant. Therefore, the quintessence mass should
be precisely of the order of $H_0$. \\
The tiny mass of the field creates problem as one loop correction
shifts the mass of the field by a huge amount $m^2 \to m^2+g M^2$(M
is cut off) unless we tune the coupling $g$ appropriately. Since
$m^2 \sim H^2_0$, the required fine tuning brings us back to
cosmological constant. Since there are no known symmetries at low
energies to control the radiative corrections, the purpose of
introducing dynamical dark energy this way stands defeated. Let us
mention an attempt to construct a string inspired axionic
quintessence for which the radiative corrections might be under
control\cite{trivedi}. However, the scenario belongs to the class of
thawing models and thereby faced
with the same level of fine tuning as cosmological constant.\\

Before we get to the next topic, we would like to comment on the
stability of fundamental scalar against radiative corrections. One
might think that the large correction to mass is the artifact of the
regularization as dimensional scheme of regularization always
involves logarithmic dependence on the cut off\footnote{MS thanks Yi
Wang for posing this question to him and he is indebted to R. Kaul
for clarifying the issue. }.In order to clarify the issue, let us
accurately compute the one loop correction to mass of the
fundamental scalar,
\begin{eqnarray}
\label{i1}
(\delta m^2)_{1~loop}\sim\int{ \frac{d^4k}{k^2+m_s^2}} \sim M^2+m_s^2 \ln M^2/m_s^2\\
\end{eqnarray}
where $m_s$ is the mass of field circulating in the loop and $M$ is
the cut off on four momentum introduced to compute the divergent
integral. It should be noticed that unlike the calculation of vacuum
energy, the cut off used here preserves Lorentz invariance.
Secondly, one often quotes the first term of (\ref{i1}) as
correction to mass which is quadratic divergent (as we did above)
when cut off is removed. Let us compute the same using dimensional
regularization,
\begin{equation}
\label{i2} ( \delta m^2)_{1~loop}\sim\int{
\frac{d^4k}{k^2+m_s^2}}\sim \frac{1}{\epsilon}+m_s^2\ln
\frac{m_s^2}{\mu^2}
\end{equation}
We notice that the first terms in both the expressions (\ref{i1})
$\&$ (\ref{i2}) are divergent and need to be subtracted; the
remaining logarithmic corrections are essentially same whether we
impose simple cut off on four momenta in the divergent integral or
we employ the dimensional regularization. We should emphasize that
it is the property of the fundamental scalar that the radiative
correction to its mass is proportional to the mass of the field it
interacts with. The dominant contribution comes from the heaviest
mass scale in the theory to which the one loop correction is
proportional to. In case there is such a mass scale in the theory,
it would destabilize the system as there is no symmetry to protect
it at low energies. This is a generic problem inherent to theories
that include  a fundamental scalar and it has nothing to do with the
regularization
 scheme we use. Indeed, the same does not happen in electrodynamics
 where the one loop correction to mass of electron is given by,
 \begin{equation}
(\delta m_e^2)_{1~loop}\sim e^2m_e^2\ln M^2/m_e^2
 \end{equation}
which is remarkable in a sense that atomic physics can rely on the
interaction of electrons and photons and can safely ignore heavier
fermions; their contribution is suppressed by inverse powers of the
corresponding heavier mass scales which is radically different from
what happens in  theory with a fundamental scalar.
\subsection{Cosmological constant, scalar field and t' Hooft criteria
of naturalness} In a healthy field theoretic set up, the higher mass
scales are expected to decouple from low energy physics. According
to t'Hooft, a parameter in the field theory is termed natural if by
switching it off in Lagrangian at the classical level enhances
symmetry of theory which is also respected at the quantum level. Let
us immediately note that cosmological constant is not a natural
parameter of Einstein theory. Indeed, in absence of matter, if we
ignore $\Lambda_b$, Eintein equations(\ref{eqeq}) admit Minkowsky
space time as solution. In this case, the underlying symmetry group,
namely, the Poincare group has 10 generators similar to the case of
de Sitter space time that one obtains as solution after invoking
cosmological constant in Einstein equations. We therefore conclude
that cosmological constant is not a natural parameter of Einstein
theory. It is also clear from the above discussion that any field
theory that contains a fundamental scalar suffers from the problem
of naturalness  see Ref.\cite{smyr}for details. In these theories a
protection mechanism should be in place. The recent discovery of
Higgs boson of mass around
 125 GeV cries for supersymmetry essential for the consistency of
 the framework.
 Clearly, both the
cosmological constant and scalar field are faced with problem of
similar nature.

Let us also emphasize that in field theory formulated in flat space
time, vacuum energy can  safely be ignored by choosing  normal
ordering. It is legitimate  as there is no known laboratory
experiment to measure the absolute value of energy; we normally
measure the difference such that the vacuum energy gets canceled in
the process. Can't we then play the following trick to address the
cosmological constant problem? Indeed, the FRW metric, is
conformally equivalent to Minkowsky space time. By a suitable
conformal transformation on Einstein-Hilbert action with
cosmological constant, we can transform to flat space time. However,
in this case, we are left with scalar field non-minimally coupled to
matter. Taking into account the fact that particle masses in the
Einstein frame become field dependent, one can demonstrate that the
scalar field in flat space time imbibes full information of FRW
dynamics. Have we then done away with cosmological constant problem?
Unfortunately, scalar field as we pointed out is plagued with the
problem of naturalness thereby one problem translates into another
equivalent one.


\section{Large scale modification of gravity and its relevance to late
time cosmic acceleration} As mentioned before, the modified theories
of gravity at large scales are essentially represented by Einstein
Gravity(GR) along with the extra degrees of freedom. For instance,
in $f(R)$ theories \cite{no,Sotiriou:2008rp,DeFelice:2010aj}, we
have one scalar degree of freedom $\varphi$ dubbed scalaron which is
mixed with the curvature in the Jordan frame. We can diagonalize the
Lagrangian by performing a conformal transformation on $f(R)$ action
 reducing the theory in Einstein frame to GR plus a scalar field
with a potential uniquely determined through $R$ and the first
derivative of f(R) with respect to R. Consistency demands that
$f'>0$ (absence of ghost) and$ f''>0$(absence of tachyonic mode or
Dolgov- Kawasaki instability). In Einstein frame, degrees of freedom
become diagonalized but $\varphi$ gets directly coupled to matter
and the coupling is typically of the order of one. We emphasize that
both the frames are not only mathematically equivalent but also
describe same physics: the relationship between physical observables
is same in both the frames. The extra degree of freedom $\varphi$
should give rise to rise to late time cosmic acceleration thereby
telling us that its mass $m_\varphi \sim H^2_0$. However,  such a
light field directly coupled to matter would grossly violate the
local physics where GR is in excellent agreement with observations .
For instance, solar physics would be safe if $m_\varphi>10^{-27}$.
It is an irony that large scale modification interferes with local
physics which is related to the fact that GR describes local physics
to a very high accuracy. Thus, if $f(R)$ to be relevant to late time
cosmic acceleration, the scalaron should appear light at large
scales and heavy locally in high density regime {\it a la } a
chameleon field \cite{Khoury:2003rn,Khoury:2003aq}. In what follows,
we shall present basic features of large scale modification of
gravity.

\subsection{Modified theories of gravity}
An important class of
modified theories can be described by generalized scalar tensor
theories. Let us for simplicity consider the following action in
Einstein frame,
\begin{equation}
\label{sta}
\mathcal{S}=\int{\sqrt{-g}d^4x\Big[\frac{M^2_p}{2}R-\frac{1}{2}(\partial
_\mu \phi)^2-V(\phi)\Big]}-\int{\sqrt{-g}d^4x
\mathcal{L}_m(\psi,A^2(\phi)g_{\mu\nu})}
\end{equation}
where $\psi$ are the matter fields and $A(\phi)$ is the conformal
coupling which relates Einstein metric $g_{\mu\nu}$ with the Jordan
metric as,
\begin{equation}
\tilde{g}_{\mu\nu}\equiv A^2g_{\mu\nu}
\end{equation}
and appears in the matter Lagrangian. We can generalize the scalar
field Lagrangian in (\ref{sta}) by including non linear higher
derivative terms dubbed galileons
\cite{Luty:2003vm,Nicolis:2004qq,Nicolis:2008in,Deffayet:2009wt},
 \cite{Deffayet:2009mn,Trodden:2011xh,Deffayet:2011gz,deRham:2010eu,Goon:2011qf,
DeFelice:2010nf,Shirai:2012iw} or generalized galieons{\it a la}
Hordenski field \cite{Horndeski:1974wa,Deffayet:2013lga}. We shall
provide  outline of galileon field dynamics in the discussion to
follow. Going ahead, we wish to point out that these fields are
central to Vainshtein screening which in turn are at the heart of
massive gravity
\cite{Fierz:1939ix,ArkaniHamed:2002sp,deRham:2010ik,deRham:2010kj,Hassan:2011hr,deRham:2011rn,Hassan:2011zd,
deRham:2010tw,Creminelli:2005qk} (for review, see
Ref.\cite{Hinterbichler:2011tt}). In the discussion to follow, we
shall first consider scalar field with potential suitable to
implement chameleon mechanism  and then turn to  massless field and
its screening using kinetic suppression.

In case of a massive field, it is instructive to write down the
equation of motion for the field in presence of the conformal
coupling by varying the action (\ref{sta}),
\begin{equation}
\Box \phi=-A'(\phi) T+\frac{dV}{d\phi}=-\frac{\alpha}{M_p}
T+\frac{dV}{d\phi};~~\alpha\equiv M_p\frac{d\ln A(\phi)}{d\phi}
\end{equation}
where $\alpha$ is  coupling constant and for simplicity, we assume
that $A(\phi)\simeq 1+\alpha \phi/M_p (\phi/M_p<<1)$. Let us note
that $f(R)$ theories correspond to $\alpha=1/\sqrt{6}$. It is
important to understand the physical meaning of $\phi$ which becomes
clear by considering the Newtonian limit in presence of the
conformal coupling. In this case, the geodesics equation is given
by,
\begin{equation}
\label{geq}
 \frac{d^2x^\mu}{d\tau}+\Gamma^\mu_{\alpha\beta}
\frac{dx^{\alpha}}{d\tau} \frac{dx^\beta}{d\tau}+\frac{\alpha}{M_p}
\partial^\mu \phi\simeq 0
\end{equation}
where the last term in the above equation is sourced by the
conformal coupling. The second term in (\ref{geq}) in Newtonian
limit yields the gradient of Newtonian potential with minus sign
supplemented by the third term due to conformal coupling,
\begin{equation}
\Phi_{tot}=\Phi_N+\frac{\alpha}{M_p}\phi
\end{equation}
We should once again remind ourselves that $\alpha$ is of the order
of one in which case the contribution of the additional term may
become comparable to $\Phi_N$. In such a scenario, the local physics
would be disturbed as the latter is described by GR with a fantastic
accuracy.  We, therefore, need to locally screen out the effects of
the extra force (fifth force) to a great accuracy which is
implemented by the chameleon mechanism for a massive field. Before
we move ahead it might be instructive to transform the action
(\ref{sta}) back to Jordan frame,
\begin{equation}
\label{bdj}
 \int{d^4x \sqrt{-g }\Big[\frac{M^2_p}{2}\Phi \tilde{R}-
\frac{ M^2_p}{2}\frac{\omega(\Phi)}{\Phi}\tilde{g}^{\alpha
\beta}\partial_\alpha \Phi
\partial_\beta \Phi-\Phi^2
V(\Phi)\Big]}+\int{d^4x\sqrt{-\tilde{g}}\mathcal{L}_m(\psi,\tilde{g}_{\mu\nu}})
\end{equation}
where $\Phi=A^{-2}(\phi)$ and $\omega(\phi)$ is given by,
\begin{equation}
\label{omega}
 \omega(\phi)=\frac{1}{2}\left[ \frac{1}{2M^2_p\left(\frac{A'}{A}
\right)^2}-3\right] \to \frac{1}{\alpha^2}=2\omega( \phi)+6.
\end{equation}
Here ``prime'' $(\ \ '\ )$ denotes derivative with respect to the
field. Let us comment on relation of Brans-Dicke parameter and the
coupling constant $\alpha$. It follows from (\ref{omega}) that
$\alpha=1/\sqrt{6}$ for $\omega=0$ which corresponds to $f(R)$ . The
coupling constant $\alpha$ as we repeatedly mentioned is typically
of the order of one whereas local gravity constrains demand that
$\omega \gtrsim 4\times 10^4$ correspondingly $\alpha$ is
vanishingly small. The latter describes the trivial regime of scalar
tensor theories and one is dealing in that case with a coupled
quintessence with negligibly small coupling. If accelerated
expansion takes place in this case, it is definitely due to flatness
of the potential. In such cases one does not need chameleon
mechanism and corresponding scalar theories are of little interest.
Let us also note that at the onset it appears from (\ref{bdj}), that
$G_{eff}=A(\phi)G$. However, what one measures in Cavendish
experiment is different and can be inferred, for instance, from weak
field limit \cite{EspositoFarese:2000ij},
\begin{equation}
 G_{eff}=G A(\phi)\left(1+2\alpha^2\right)
\end{equation}
where the expression in parenthesis is due to the exchange of the
scalaron.\\

It is clear from the aforesaid that chameleon is essential for
generic modified theories. In what follows we outline the underlying
concept of chameleon screening.

\begin{figure}[h]
\includegraphics[scale=0.4
]{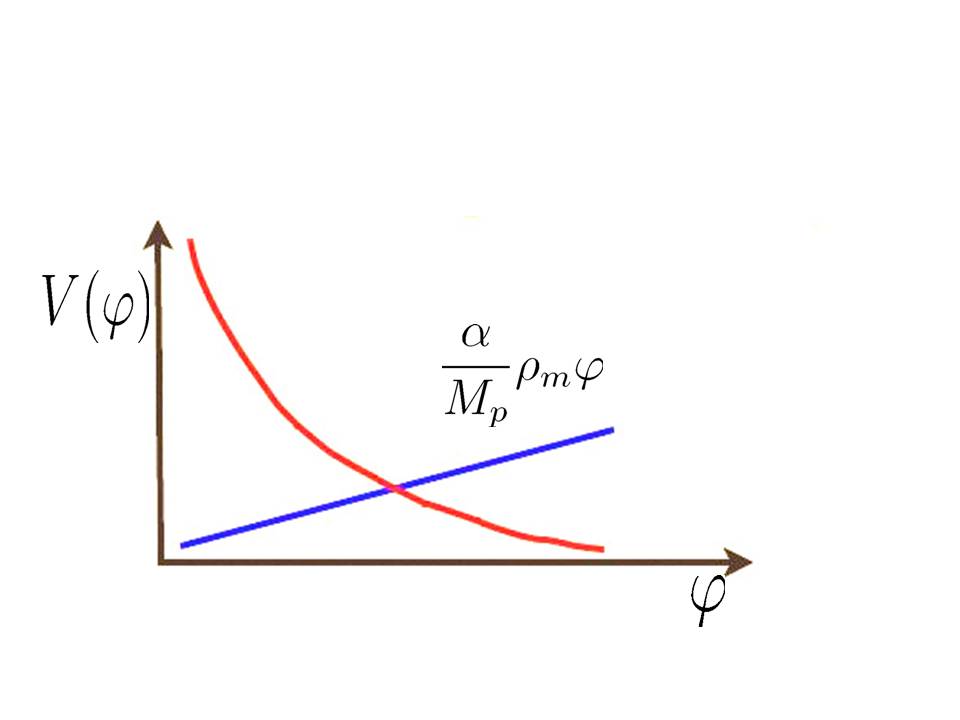} \caption{ Effective potential for a chameleon field.
$V(\phi)$ is generically a run away potential without a minimum. The
effect of direct coupling with matter modifies the potential such
that the effective potential acquires minimum. Higher is the density
of environment, closer would be the minimum to the origin. The
potential is a monotonically decreasing function(we might imagine,
($V \sim 1/\phi^n$) such that its second derivative $V''(\phi)$ is
also monotonically decreasing and positive. } \label{chp}
\end{figure}

\subsection{Chameleon theories: basic idea}
In order to set the basic notions of chameleon screening, let us
first for simplicity consider a massive scalar field non-minimally
coupled to matter \cite{Amendola:1999er},
\begin{equation}
\label{lch} \mathcal{ L}=-\frac{1}{2}\partial_\mu \varphi
\partial^\mu \varphi-\frac{1}{2}m^2\varphi^2+\frac{\alpha} {M_p}\varphi T
\end{equation}
which on varying with respect to $\varphi$ gives the following
equation of motion,
\begin{equation}
(\Box+m^2)\varphi=-\frac{\alpha}{M_p}T
\end{equation}
 In this case, for a given static
point source of mass $M$,
 $r=0,~~T=-M\delta^3(r)$, the potential sourced by the field is
 given by,
\begin{equation}
\frac{\alpha}{M_P}\varphi=-{2\alpha^2 G M}\frac{ e^{-mr}}{ r}
\end{equation}
which is the extra contribution to the gravitational potential of
the  point source  due to scalaron. The total potential is then
given by,
\begin{equation}
\Phi_{tot}=-\frac{GM}{r}\left(1+2\alpha^2e^{-mr}\right)
\end{equation}

 As mentioned before, $\alpha$ is typically of the order of
one. Hence the extra force mediated by the exchange of scalaron
between two point masses is of the order of the gravitational force
for light mass $mr<<1$, relevant to late time cosmic acceleration.
The latter is equivalent to $G\to G_{eff}=G(1+2\alpha^2)$ which is
clearly in conflict with local physics. The consistency at the level
of solar system demands that $mr_{AU}<<1$ or $m>> 10^{-27}GeV$. It
is therefore clear that the mass of scalaron should be environment
dependent $m(\rho)$ $-$ light in low density regime(at large scales)
and heavy in high density regime locally. We shall briefly
demonstrate in the discussion to follow how the chameleon field
generated by an extended massive source may get effectively
decoupled
from the source leaving local physics intact.\\
\subsection{Chameleon at work}
\begin{figure}[h]
\includegraphics[scale=0.4]{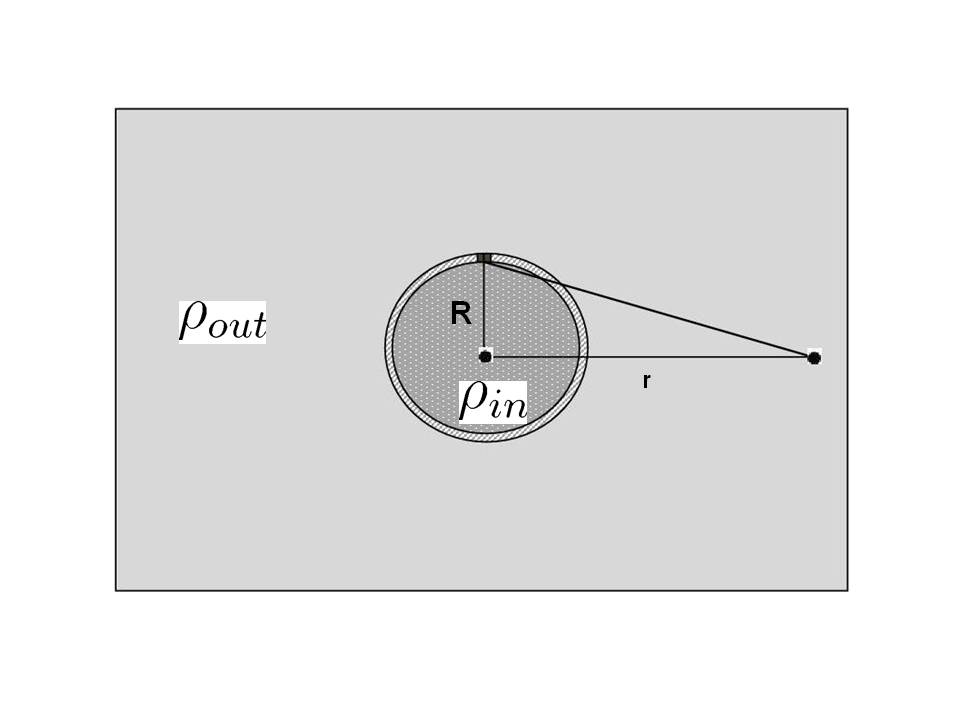}
\caption{Figure shows a body of mass ${M}$ with a density
$\rho_{in}$ embedded in an environment with density
$\rho_{out}<<\rho_{in}$. Contribution to the field profile at
distance $r$ from the massive body comes from a thin layer under the
surface of the massive body due to Yukawa suppression in the
interior. }
 \label{thins}
\end{figure}

Let us briefly examine how the chameleon mechanism operates
\cite{Khoury:2003rn,Khoury:2003aq}. The
aforesaid discussion makes it clear that we should choose a suitable
scalar field potential to achieve the goal. The inverse power law
potentials are generic, they become shallow at late time and might
give rise to late time acceleration. The effective potential in
presence of the coupling is given by,
\begin{equation}
V_{eff}=V(\varphi)+\frac{\alpha}{M_p}\rho_m \varphi
\end{equation}
It is clear from Fig. \ref{chp} that $V_{eff}$ has a minimum which
is closer to the origin higher is the density of the environment.
Since $V''(\phi)$ is positive and monotonously decreasing for the
generic cases, the mass of the field around the minimum is larger
higher is the matter density of
the environment and vice versa what was sought for .\\
We next need to compute the field profile for an extended body of
mass $\mathcal{M}$. In case of the gravitational potential {\it a
la} Newton, the answer is simple: the point particle mass in the
expression of its gravitational potential gets replace by
$\mathcal{M}$. It should be emphasized that such a privilege is
restricted to $1/r$ potential only. In any other case and in
particular in the case under consideration, the potential of an
extended body, apart from its mass would also depend upon its
density. The contribution to the field profile coming from the
interior gets Yukawa suppressed due to its large mass in high
density regime. Contribution, if any, comes from a thin layer under
the surface of the body, see Fig.\ref{thins}.

As shown in Refs. \cite{Khoury:2003rn,Khoury:2003aq},
\begin{equation}
\label{thin}
 \frac{\alpha}{M_P} \varphi=-\frac{GM}{r}\alpha^2
\epsilon_{thin}
\end{equation}
where $ \epsilon_{thin}$, the thin shell parameter is given by
\begin{equation}
\epsilon_{thin}\propto
\frac{\varphi_{min}^{out}(\rho)-\varphi_{min}^{in}(\rho)}{\Phi_\mathcal{M}}
\end{equation}
where ${\Phi_\mathcal{M}}$ is the Newtonian potential of the
extended body. Since,
$\varphi_{min}^{in}(\rho)<<\varphi_{min}^{out}(\rho)$ because of the
high density inside the body and thus can be dropped. The success of
chameleon mechanism then depends upon the fact that the
gravitational potential for an extended body, say Sun, is large and
$\varphi_{min}^{out}(\rho)$ is small in the solar system. As for the
accuracy of GR, the agreement can be reached by suitably choosing
model parameters through $\varphi_{min}^{out}(\rho)$. As a result,
the effective coupling, $\alpha_{eff}=\alpha \epsilon_{thin}$ in
Eq.(\ref{thin}) can be made as small as desired thereby effectively
giving rise to  decoupling of the field from
the source or the screening of the extra force.\\
At the onset, it looks like that we have succeeded in getting late
time cosmic acceleration via the extra degree of freedom $\varphi$,
which imbibes large scale modification of gravity, keeping it
invisible locally. However, a close scrutiny of chameleon theories
reveals that required screening of extra degree(s) leaves no scope
of these theories for late time cosmic acceleration. The problem
stems from high accuracy of Einstein theory in solar system and
laboratory experiments.
\begin{figure}[h]
\includegraphics[scale=0.5]{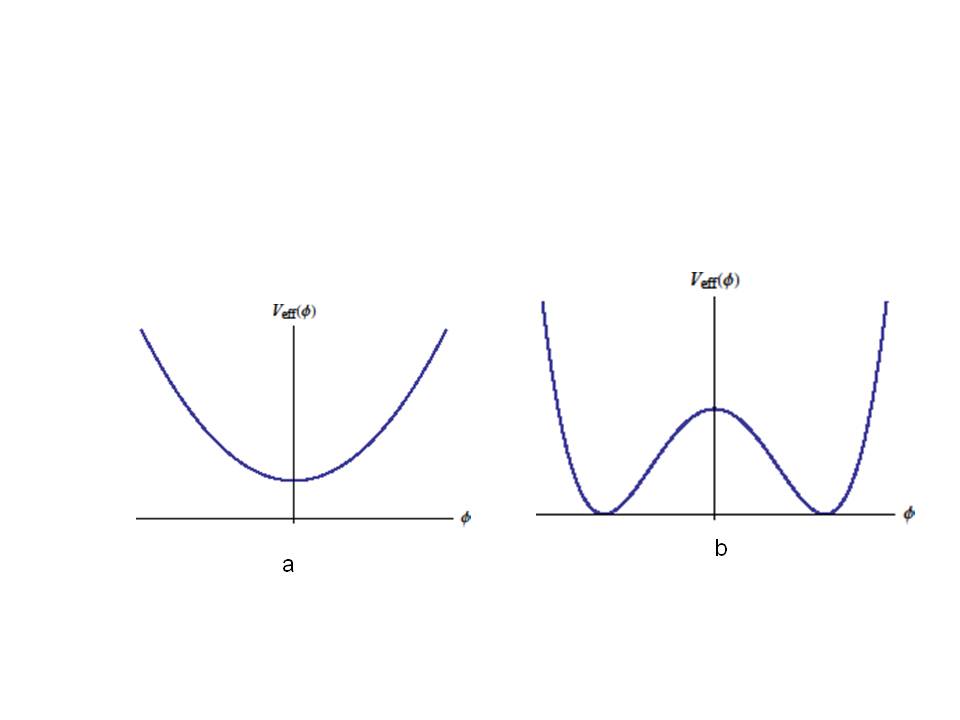}
\caption{ Fig.\ref{sym}(a) displays the symmetron potential in high
density regime. In this case the system resides in the symmetric
vacuum $\phi=0$. On the other hand, in the low density regime around
$\rho=\rho_{cr}$, the symmetric state is no longer a true ground
state  (Fig.\ref{sym}(b)), the system then makes transition to one
of the ground states giving rise to spontaneous symmetry breaking of
$Z_2$ symmetry of the underlying system} \label{sym}
\end{figure}
\section{Spontaneous symmetry breaking in cosmos: A beautiful idea that does not work}
As mentioned before, universe has undergone a transition from
deceleration to acceleration between $z=0$ and $z=1$. It is tempting
to relate the latter to breaking of a hypothetical symmetry which
can  be realized by invoking a specific conformal coupling
\cite{Hinterbichler:2010es,Hinterbichler:2011ca,Bamba:2012yf}.
 Let us very briefly out line the basic features of the the model
 dubbed
symmetron which is based upon the following Einstein frame action
\begin{align}
\label{eq:action} \mathcal{S}=\int
d^4x\sqrt{-g}\Bigl[\frac{M^2_p}{2}R-\frac{1}{2}\partial_\mu \phi
\partial^\mu \phi-\frac{\mu^2}{2}\phi^2 -
\frac{\lambda}{4}\phi^4\Bigr]
+\mathcal{S}_m\Bigl[A^2(\phi)g_{\mu\nu},\Psi_m\Bigr]
\end{align}
The symmetron potential is invariant under $Z_2$ symmetry ($\phi \to
-\phi$) and one can preserve this symmetry in the effective
potential  by making the following choice for $A(\phi)$
\cite{Hinterbichler:2010es,Hinterbichler:2011ca}
\begin{equation}
A(\phi)=1+\frac{\phi^2}{2M^2}~(\phi<<M)
\end{equation}
where $M$ is a mass scale in the model. The effective potential then
takes the following form
\begin{equation}
V_{eff}=\frac{1}{2}\left(\frac{\rho}{M^2}-\mu^2\right)\phi^2+\frac{\lambda}{4}\phi^4
\end{equation}
The mass of the field now depends upon the density of environment,
naively, the field mass is given by, $m^2_{eff}={\rho}/{M^2}-\mu^2$.
Thus in high density regime, mass depends upon density linearly,
$m^2_{eff}\sim \rho/M^2>0$. In this case, the system resides in the
symmetric vacuum specified by $\phi=0$. The requirement of local
gravity constraints  puts an upper bound on the mass scale, $M$ and
there is no reason for it to be consistent  with dark energy. We
should note that in case of chameleon, there is more flexibility,
the mass depends on density
non-linearly. As shown in Ref.\cite{Hinterbichler:2010es,Hinterbichler:2011ca}, $M \leq 10^{-4}M_{p}$. \\
As the density redshifts with expansion and $\rho$ drops below
$\mu^2/M^2$, tachyonic instability builds in the system and the
symmetric state $\phi_0=0$ is no longer a true minimum. The true
minima are then given by(see Fig\ref{sym})
\begin{equation}
\phi_0=\pm \sqrt{\frac{\mu^2-\rho/M^2}{\lambda}}.
\end{equation}
The mass of the symmetron field about the true minimum is given by,
$m_{s}=\sqrt{2}\mu$. Universe goes through a crucial transition when
late time acceleration sets in around the redshift $z\sim 1$. It is
therefore natural  to assume that the phase transition or symmetry
breaking takes place when $\rho\sim\rho_{cr}$. Hence we conclude
that,
\begin{equation}
\rho_{cr} \simeq M^2\mu^2 \to \mu^2 \simeq \frac{H_0^2
M^2_{pl}}{M^2} \to m_{s}\simeq \frac{H_0 M_{p}}{M}
\end{equation}
This means that $m_{s}\ge 10^4H_0$ which is larger than the required
quintessence mass by several orders of magnitude. In this case, the
field rolls too fast around the present epoch making itself
untenable for cosmic acceleration. Invoking the more complicated
potential with minimum with the required height does not solve the
problem. In this case field would continue oscillating around the
minimum for a long time and would not settle in the minimum unless
one arranges symmetry breaking very near to $z=0$ by invoking
unnatural fine tuning of parameters. There is no doubt that
symmetron presents a beautiful idea but, unfortunately, fails to be
relevant to late time cosmic acceleration. We believe that it would
find a meaningful application in cosmology in some other form.

\subsection{ Scope of chameleon for late time cosmic acceleration}
The large scale modification of gravity effects the gravitational
interaction because of the two reasons. (1) The exchange of extra
degree(s) of freedom which couples with  matter source roughly with
the same strength as graviton and whose local influence needs to be
screened using a suitable mechanism. (2) The conformal coupling
$A(\phi)$ also modifies the strength of gravitational interaction.
And to pass the local tests, $A(\phi)$ should be very closely equal
to one in high density regime in  chameleon supported theories. The
transition Universe has undergone during $0<z<1$, is a large scale
phenomenon and one might think that the mass screening which is a
local effect should not impose severe constraints on how $A(\phi)$
changes during the period acceleration sets in. It turns out that
the change the conformal coupling suffers as redshift changes from
one to zero is negligibly small. Then the question arises, can such
a conformal coupling be relevant to late time acceleration? \\
It is well known that the de Sitter Universe is conformally
equivalent to the Minkowski space-time. Does the conformal
transformation changes physics? By `physics', we mean the
relationship between physical observables. In the Einstein frame we
have the Minkowski space-time where there is a scalar field sourced
by the conformal coupling which directly couples to matter. The
masses of all material particles are time dependent by virtue of the
conformal coupling $A(\phi)$. Consequently, one would see the same
relations between physical observables in both the
frames\cite{misao}. The acceleration dubbed {\it self acceleration}
is the one which can be removed (caused) by conformal coupling
\cite{scope}. Late time cosmic acceleration which is not related to
conformal coupling is caused by the  slowly rolling (coupled)
quintessence and is not a generic effect of modified theory of
gravity. Indeed, this is the case if we adhere to chameleon
screening. In what follows we shall describe how it happens. We have
the following relation between scale factors in Einstein and Jordan
frames,
\begin{equation}
\label{EJ1} a^J(t^J)=A(\phi)a^E(t^E)\, ,\quad dt^J=A(\phi)dt^E\, ,
\end{equation}
and as for the conformal time,$dt=a(t)d\eta$, it  is same in both
the frames . In (\ref{EJ1}), $a^J$( $(a^E)$) denote scale factor and
$t^J$( $(t^E)$) the cosmic time in the Jordan (Einstein) frame.

Let us take the derivatives with respect to the Jordan cosmic time
$t^J$ of $a^J(t_J)=A(\phi)a^E(t^E)$ left right,
\begin{equation} \dot{
{a}}^J(t^J)=\frac{1}{A}\frac{d}{dt^E}\left(Aa^E\right)
\end{equation}
where derivative of Einstein frame quantities is taken with respect
Einstein frame time. Differentiating the last equation again with
respect to Jordan time gives,
\begin{equation}
\label{EJ2} \ddot{a}^J(t^J)=\frac{1}{A}\left(\ddot{a}^E
+\frac{\ddot{A}}{A}a^E-\frac{\dot{A}^2}{A^2}a^E
+\frac{\dot{A}}{A}\dot{a}^E\right)\, .
\end{equation}
By time derivative ``dot'' $(\ \dot\ \ )$ of quantity in the Jordan
(Einstein) frame, we mean time derivative with respect to the Jordan
(Einstein) time $t^J$ ($t^E$). Multiplying this equation left right
by $a^J=Aa^E$, we have,
\begin{equation}
\label{EJ3} \ddot{a}^J
a^J-\ddot{a}^Ea^E=\left(\frac{\ddot{A}}{A}-\frac{\dot{A}^2}{A^2}\right){(a^E)}^2
+\frac{\dot{A}}{A}\dot{a}^Ea^E \, .
\end{equation}
The right hand side can be put in compact form by changing the
Einstein frame time to the conformal time from $\left(d/dt^E \to
\left(1/a\right) \left(d/d\eta\right)\right)$.

Indeed, following Ref.~\cite{scope}, we have a relation which
relates $\ddot{a}$ in both the frames,
\begin{equation}
\label{EJ5} \ddot{a}^Ja^J-\ddot{a}^Ea^E
=\left(\frac{A''}{A}-\frac{A'^2}{A^2}\right)
=\left(\frac{A'}{A}\right)'\, .
\end{equation}
where ``prime'' $(\ \ '\ )$ denotes the derivative with respect to
conformal time in the Einstein frame. Let us note that acceleration
in the Einstein frame cannot be caused by conformal coupling,
\begin{equation}
\label{EJ6} \frac{\ddot{a}}{a}=-\frac{1}{6M_\mathrm{Pl}^2}
\left((\rho_\phi+3P_\phi)+\alpha\rho A(\phi)\right)\, .
\end{equation}
Thus in case acceleration takes place in the Einstein frame, it can
only be caused by slowly rolling quintessence
($\rho_\phi+3P_\phi<0$). This implies that acceleration in the
Jordan frame and no acceleration in the Einstein frame is generic
effect of conformal coupling or large scale modification of gravity.
In this case, while passing from the Jordan to the Einstein frame,
acceleration is completely removed.  We can adopt the following
definition \cite{scope},
\begin{equation}
self~ acceleration:~~~~~ \ddot{a}^Ea^E<0~;~~~ \ddot{a}^Ja^J>0
\end{equation}
  which implies
\begin{equation}
\label{EJ7} \left(\frac{A'}{A}\right)\ge \ddot{a}^Ja^J\, .
\end{equation}
Next, we can express $A'$  through its variation
  over one Hubble
(Jordan) time). It then follows that
\begin{eqnarray}
&&A'=\dot{a}^J \Delta A; ~~~\Delta
A=\left(\frac{1}{H^J}\frac{dA}{dt^J}\right)\\
&&  \frac{d}{dt^J}\left(\dot{a}^J \frac{\Delta A}{A}\right)\ge
\ddot{a}^J\
\end{eqnarray}

Integrating  the above relation left right, we find \cite{scope}
\begin{equation}
\label{EJ9} \frac{\Delta{A}}{A}\gtrsim 1\, .
\end{equation}

As demonstrated in Ref.~\cite{scope}, screening imposes a severe
constraint on the change of coupling during the last Hubble time,
$\Delta A \ll 1$. Thus self acceleration cannot take place in this
case. In most of the models supported by chameleon screening,
acceleration takes place in both frames such that $\ddot{a}^Ja^J$
and $\ddot{a}^Ea^E$ cancel each other with good accuracy or $\Delta
A \ll 1$. In this case acceleration can only be caused by slowly
rolling quintessence. \\
We therefore conclude that theories of large scale modification
based upon chameleon screening have no scope for late time cosmic
acceleration. These theories are also plagued with the problem of
large quantum corrections due to the large mass of the chameleon
field required to satisfy the local gravity constraints.
\begin{figure}[h]
\includegraphics[scale=0.5
]{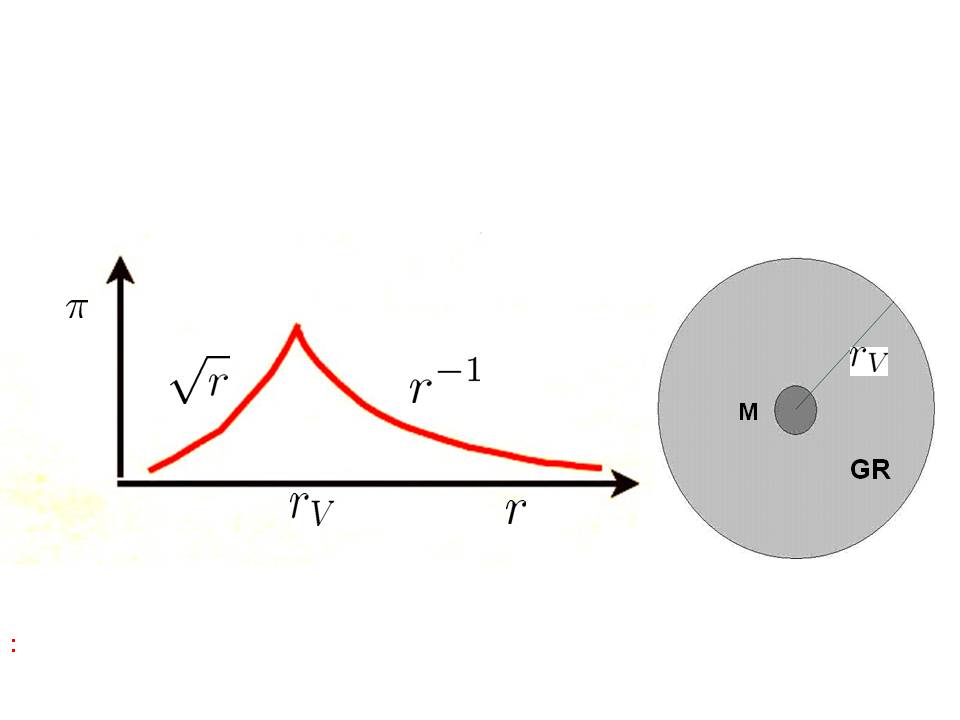} \caption{The figure shows Vainshtein mass screening
characterized by galileon field profile for massive body of mass
${M}$. The character of non linearity crucially changes the $\pi$
profile around the body below Vainshtein radius $r_V$. Within this
radius, the $\pi$ mediated force is negligible as compared to the
Newtonian force and GR is left intact there. The modification can be
felt beyond Vainshtein radius only.} \label{vp}
\end{figure}
\subsection{Modified theories of gravity: Vainshtein screening}
It is clear from the above discussion that chameleon mechanism would
fail if the mass of the field is zero. How then to screen the local
effects induced by such a field? There is superior field theoretic
mechanism for hiding the massless degrees of freedom know as
Vainshtein mechanism \cite{Vainshtein:1972sx}. It does not rely on
mass of the field and operates dynamically through kinetic
suppression which was suggested by A. Vainshtein in 1972 to address
the problem of vDVZ \cite{vanDam:1970vg,Zakharov:1970cc}
discontinuity in Pauli-Fierz theory \cite{Fierz:1939ix}. This
mechanism can be consistently implemented through galileon field
$\pi$
\cite{Nicolis:2008in,Babichev:2013usa,Babichev:2009us,Gannouji:2010au,Ali:2010gr,Chow:2009fm,Ali:2012cv}
whose Lagrangian apart from the standard kinetic term contains
non-linear derivative terms of specific form. The strong
non-linearities become active around a massive body below Vainshtein
radius which effectively decouple the field from the source leaving
GR intact there. In a space time of dimension $n$, there is a fixed
number of total derivatives one can construct using
$\partial_\mu\partial_\nu \pi$ correspondingly there is fixed number
of galileon Lagrangians in each space time dimension.\\

Let us list the galileon Lagrangians in case of four dimensions \cite{Nicolis:2008in},
\begin{eqnarray}
&&\mathcal{L}_1=\pi \\
&&\mathcal{L}_2=-\frac{1}{2}(\partial_\mu \pi)^2\\
&&\mathcal{L}_3=-\frac{1}{2}(\partial_\mu \pi)^2\Box \pi\\
&&\mathcal{L}_4=-\frac{1}{2}(\partial_\mu \pi)^2\left[(\Box \pi)^2-
\partial_\mu\partial_\nu\pi \partial^\mu\partial^\nu\pi\right]\\
&& \mathcal{L}_5=-\frac{1}{2}(\partial_\mu \pi)^2\left[(\Box \pi)^3-
3\Box \pi(\partial_\mu\partial_\nu\pi \partial^\mu\partial^\nu\pi)+
2\partial_\alpha\partial_\beta\pi \partial^\beta\partial^\delta\pi
\partial_\alpha\partial^\delta\pi
 \right]
\end{eqnarray}
Due to the specific underlying structure from which the galileon
Lagrangians can be constructed, the equations of motion for galileon
field are of second order despite the higher derivative terms in the
Lagrangian \cite{Nicolis:2008in}. Secondly, the galileon Lagrangians
are invariant under shift symmetry, $\pi\to \pi+b_\mu x^\mu+c$, in
flat space time thank to which their equations of motion for can be
represented as the divergence of a conserved current corresponding
to the shift symmetry. Before we proceed ahead, let us remark that
physics of Vainshtein mechanism is already contained in the lowest
order Lagrangian $\mathcal{L}_3$ \cite{Chow:2009fm,Ali:2012cv};
higher order Lagrangians add nothing to it.  However,
$\mathcal{L}_3$ alone can not give rise to de Sitter solution needed
for late time cosmology; we need at least $\mathcal{L}_4$ to serve
the purpose \cite{Gannouji:2010au,Ali:2010gr}. Since we will not
address the phenomenological issues of galileon field applied to
late time cosmology, we shall restrict ourselves to the lowest order
galileons.

\subsection{Vainshtein mechanism: Basic idea}
In case of the chameleon, the mass screening relied on the effective
potential \cite{Khoury:2003rn,Khoury:2003aq},
\begin{equation}
\Box \varphi=\left(V(\varphi)+\frac{\alpha}{M_p}\rho_m \varphi
\right)_{,\varphi}
\end{equation}
such that the mass of the field turned large in high density regime
which then decouples it from the source. In case of massless field,

\begin{equation}
\label{zmf} \Box \varphi=+\frac{\alpha}{M_p}\rho_m
\end{equation}
chameleon ceases to work.  We observe that the multiplication of the
left hand side of (\ref{zmf}) by a constant is equivalent to
dividing the coupling constant $\alpha$ on the right hand side by
the same constant. The latter means that enhancement of kinetic term
effectively suppresses the coupling of the field to matter. However,
we can not do it by hand, it should be implemented by field
theoretic framework. In Vainshtein mechanism, the latter is achieved
dynamically in a very intelligent manner by
making use of the galileon field.\\
Let us briefly illustrate how kinetic suppression takes place in
galileon field theory. To this end as mentioned before, it is
sufficient to consider the lowest galileon Lagrangian
$\mathcal{L}_3$ which gives rise to the following equation of motion
\cite{Nicolis:2008in,Chow:2009fm,deser,Ali:2012cv},
\begin{equation}
\label{geq} \Box \pi+\frac{1}{\Lambda^3}[(\Box \pi)^2-\partial^\mu
\partial^\nu \pi\partial_\mu \partial_\nu \pi]\nonumber\\
=-\frac{\alpha}{M_p}T
\end{equation}
where $\Lambda=(m^2M_p)^{1/3}$ is the cut off in the effective
Lagrangian and $m \sim H_0$. The second term on the left is
non-linear which may dominate over the standard kinetic term at
small scales. Indeed, for a static source of mass
M($T=-M\delta^3(\bf{r})$), in case of spherical symmetric solution
of interest to us, Eq.(\ref{geq}) acquires the following form,
\begin{equation}
\frac{1}{r^2}\frac{d}{d r}\left(r^3\left[(\pi
'/r)+\frac{1}{\Lambda^3}(\pi'/r)^2\right]\right)=\frac{\alpha}{
M_p}M\delta^3({r})
\end{equation}
which thank to the total derivative structure of equation of motion
readily integrates to
\begin{equation}
\label{pi1}
\left(\frac{\pi'(r)}{r}\right)+\frac{1}{\Lambda^3}\left(\frac{\pi'(r)}{r}\right)^2=\alpha\frac{r_s}{r^3}
\end{equation}
where $r_S$ is the Schwarzschild radius of the massive body.
 We observe that at small distance the second term in the
expression (\ref{pi1}) dominates over the first which tells us that
\begin{equation}
\pi'=\left(\frac{r_S \alpha m^2}{r}\right)^{1/2}
\end{equation}
 As a result
the extra force due to galileon field is suppressed as compared to
the gravitational force in the neighborhood of the massive body(see
Fig.\ref{vp}),
\begin{equation}
\frac{F_\pi}{F_{\text{grav}}}=\left(\frac{r}{r_V}\right)^{3/2}<<1,~~~~r<<r_V
\end{equation}
 where the Vaishtein radius, is given by,
\begin{equation}
r_V=\left(\frac{r_S}{m^2\alpha} \right)^{1/3}
\end{equation}
On the other hand, at large scales the usual kinetic term dominates
over the nonlinear term  and Galileon force becomes comparable to
gravitational force,
\begin{equation}
\pi'=\frac{{r_S \alpha}}{r^2}~~~\Rightarrow
\frac{F_\pi}{F_{\text{grav}}}\sim 1
\end{equation}
Let us estimate $r_V$ for Sun,
\begin{equation}
r_V=\frac{GM_s}{m^2}=\frac{M_s}{H^2_0 M^2_p}\simeq 100 pc
\end{equation}
Hence, solar physics will not feel the presence of galileon field;
any modification of gravity  due the galileon degree of freedom is
locally screened out due to kinetic suppression leaving GR intact in
a radius much larger than the solar dimensions. For our galaxy,
$r_V\simeq 1.2$ Mpc; the effect of galileon field might be felt at
large distance through late time cosmic acceleration. It is
worthwhile to note that galileon field is stable under quantum
corrections unlike the chameleon.\\
\subsection{Galileons and their higher dimensional descendants}
Galileon field provides with a well defined field theory in 4
dimensions which is ghost free. On the other hand we have well
defined and consistent extension of Einstein gravity in higher
dimensions. In five and six dimensions, the Einstein-Hilbert action
is extended by including the Gauss-Bonnet term
\cite{Dadhich:2005mw}, in further higher dimensions, the Lovelock
structure comes into play \cite{Lovelock:1971yv,Lovelock:1972vz}. In
fact, Gauss-Bonnet term is the simplest form of Lovelock Lagrangian.
Thus, in each space time dimension, the consistent gravity action
which leads to second order equations of motion thereby free from
Ostrogradki ghosts \cite{Ostrogradski}, is fixed. It is tempting to
think that the two ghost free systems, the Galileon field theory in
four dimensions and higher dimensional Lovelock gravity, are some
way related to each other. In fact the galileon field theory in four
space time dimensions is a representative of higher dimensional
gravity {\it a la } Lovelock. It is interesting that dimensional
reduction of $R+\alpha R^2_{GB}$ gives rise to lower order galileon
Lagrangian, $L_3$, the roll of galileon field is played by the
dilaton field. In what follows we briefly out line how this
connection between
two ghost free theories is established.\\

Let us consider five dimensional gravity where Einstein-Hilbert is
supplemented with Gauss-Bonnet term,
\begin{equation}
\mathcal{S}=\int{d^5x \sqrt{-g^5}\left(R+\alpha R^2_{GB}\right)}
\end{equation}
which is the simplest form of Lovelock theory. We then use the
standard prescription to reduce the action to four dimension and use
the following metric ansatz,

\begin{equation}
ds^2=g_{\mu\nu}dx^\mu dx^\nu+e^\pi (dx^5)^2
\end{equation}
where the scalar field $\pi$ appearing in the metric plays the role
of the size of the extra dimensions. The dimensional reduction
assuming the extra dimension to be compact, gives the following
action,
\begin{equation}
\mathcal{S}=\int{dx^4\sqrt{-g}e^{\pi/2}\left(R+d_1(\partial_\mu\pi)^2
+ \alpha\left(R^2_{GB}+d_2G_{\mu\nu}\partial^\mu \pi\partial^\nu \pi
+ d_3(\partial_\mu
\pi)^4+d_4(\partial_\mu\pi)^2\Box\pi\right)\right)}
\end{equation}
The last term in the reduced action is the lowest order galileon
term $L_3$. It is then tempting to go beyond Gauss-Bonnet, including
higher order  Lovelock terms. In this case, it was demonstrated in
Ref\cite{LG}. that the dimensional reduction reproduces higher order
Galileon Lagrangians. It is therefore not surprising that galileon
field theory in four space time dimensions is ghost free$-$
Galileons are the representatives of higher dimensional
Lovelock theory in four dimensions.\\
\section{Glimpses of massive gravity}
It is commonly believed  that an elementary particle of mass $m$ and
spin $s$ is described by a field which transforms according to a
particular representation of Poincare group. In field theory,
formulated in flat space time, mass can either be introduced by hand
or generated through spontaneous symmetry breaking but general
theory of relativity is not formulated as a field theory. One could
naively consider the metric $g_{\mu\nu}$ as field and try to
introduce mass via the invariants, $detg_{\mu\nu}$ or $Tr
g_{\mu\nu}$ which obviously do not serve the purpose. Hence we
require a field which in some sense could represent gravity. The
spin-2 field $h_{\mu\nu}$ should be relevant to gravity as it shares
an important property of universality with Einstein general
relativity {\it a la} Weinberg theorem . It  states that the
consistent quantum field theory of a spin$-$ 2 field in Minkowski
space time is possible provided the field interacts with all  other
fields including itself with the same coupling. General theory of
relativity can be thought of as an interacting theory of
$h_{\mu\nu}$ field. It is therefore natural to first formulate the
field theory of massive spin 2 field in flat space time and then
extend it to
non-linear background.\\

Before we proceed further, let us remember, how objects with spin-0,
spin-1 and spin-2 transform under Lorentz transformation
$\Lambda_{\mu\nu}$,
\begin{eqnarray}
&&\text{ spin-0}~~~~~~~~~~~\phi'=\phi\\
&&\text{spin-1}~~~~~~~~~~~~~A'_{\mu}=\Lambda_\mu^{\alpha}A_{\alpha}\\
&&\text{spin-2}~~~~~~~~~~~~~h_{\mu\nu}=\Lambda^\alpha_\mu
\Lambda^\beta_\nu h_{\alpha\beta}.
\end{eqnarray}

 The linear massive theory of gravity of $h_{\mu\nu}$ was
formulated by Fierz and Pauli  in  1939 \cite{Fierz:1939ix} with a
motivation to write down the consistent relativistic equations for
higher spin fields including spin-2 field. Let us first cast the
relativistic equations of spin-0 and spin-1 fields,
\begin{eqnarray}
&&(\Box+m^2)\phi=0\\
&&(\Box + m^2)A_{\mu}=0;~ \partial_\mu A^{\mu}=0
\end{eqnarray}
It is important to note that the condition $\partial _\mu A^\mu=0$
is in built in the equation of motion and not imposed from outside.
Indeed, from the Lagrangian of massive vector field
\begin{equation}
\mathcal{L}=-\frac{1}{4}F_{\mu\nu}F^{\mu\nu}-\frac{1}{2}m^2A_\mu
A^\nu
\end{equation}
follows the following equations of motion,
\begin{equation}
\partial_\mu F^{\mu\nu}+m^2 A^\nu=0
\end{equation}
which upon taking the divergence on both sides immediately gives us
$\partial_\mu A^\mu=0$. Thus this condition for massive vector field
follows from the equations of motion themselves. Massive vector
field has clearly three degrees of freedom. It is important to
notice that this condition can no longer be derived from the
equation of motion in the $m \to 0 $ limit which is consistent with
the fact that we have gauge invariance in this case which allows us
to get rid of two un physical degrees of freedom. Gauge invariance
allows to fix the gauge which can be done in infinitely many ways.
For instance we can choose the radiation gauge, $A_0=0$ and ${\bf
\nabla}.{\bf A}$ leaving behind two transverse degrees of freedom.

 Respecting relativistic invariance, we could also choose Lorentz gauge,
 $\partial_\mu A^\mu=0$.  In massless case, this condition is imposed from out side
in view of gauge freedom and this  should clearly be distinguished
from $\partial_\mu A^\mu=0$ occurring in case of massive vector
field as a consequence of equations of motion. Lorentz gauge does
not completely fix the gauge invariance. Indeed, there is a residual
gauge invariance, namely, $A_\mu \to A_\mu+\partial_\mu \alpha$ such
that $\Box \alpha=0$ which when fixed leaves behind two physical
degrees of freedom.

Let us now cast the equation of motion of $h_{\mu\nu}$,
\begin{equation}
\label{heq}
 (\Box+m^2)h_{\mu\nu}=0;~~\partial^\mu
h_{\mu\nu}=0;~~h_\mu^\mu\equiv h=0
\end{equation}
which tells us that massive graviton in Pauli-Fierz (PF) theory has
five degrees of freedom \cite{Hinterbichler:2011tt}. In accordance
to our expectations, the number of degrees of freedom, $2s+1$ is 3
for massive vector field and 5 for massive graviton. The first
condition on $h_{\mu\nu}$ is analogous to the case of vector field.
The vanishing of trace of $h_{\mu\nu}$ is very specific to linear
theory and we will come back to this point later in our discussion.
 The equations of motion (\ref{heq}) can be obtained from  PF Lagrangian which has the following
form,
\begin{eqnarray}
\label{pf}
&&\mathcal{L}_{PF}=\mathcal{L}_{m=0}-\frac{1}{2}m^2\left(h_{\mu\nu}h^{\mu\nu}-h^2\right)\\
&&\mathcal{L}_{m=0}=\frac{1}{2} \partial_\lambda h_{\mu\nu}
\partial^\lambda h^{\mu\nu}+\partial_{\mu} h_{\nu\lambda}
\partial^\nu h^{\mu\lambda}-\partial_\mu h^{\mu\nu} \partial_\nu
h+\frac{1}{2}
\partial_\lambda h \partial^\lambda h
\end{eqnarray}
The first term in (\ref{pf}) describes the massless graviton and can
be obtained by considering small perturbations around flat space
time, $g_{\mu\nu} \to g_{\mu\nu}+h_{\mu\nu}$ and expanding the
Einstein-Hilbert Lagrangian in $h_{\mu\nu}$ up to quadratic order.
It is easy to verify that the massless Lagrangian, is invariant
under the following gauge transformation,
\begin{equation}
\label{gin} h_{\mu\nu}\to
h_{\mu\nu}+\partial_\mu\xi_\nu+\partial_\nu\xi_\mu
\end{equation}
which fixes the relative numerical values of coefficients in
$\mathcal{L}_{m=0}$. The second term in (\ref{pf}) is the
Pauli-Fierz mass term which breaks the gauge invariance (\ref{gin})
\cite{Hinterbichler:2011tt}. The PF mass term includes two
invariants that one can form using the spin-2 field. Let us notice
that the mass term could in general be a linear combination of these
invariants, $ c_1 h_{\mu\nu}h^{\mu\nu}-c_2h^2$; one of the
multiplicative constant, say, $c_1$ could be absorbed in $m^2$
leaving the other one $c_2/c_1$, arbitrary. The PF mass terms
corresponds to an intelligent tuning of the coefficient which
excludes the ghost from linear theory, we shall come back to this
point in the forthcoming discussion.

Recently, there was an upsurge of interests in massive gravity. A
ghost free generalization of Pauli-Fierz to non linear background
known as dRGT was discovered by de Rham, Gabadadze and Tolley \cite{deRham:2010ik,deRham:2010kj}.
However, the motivation to go for massive gravity now is quite
different from the original one. Adding mass to graviton might
account for late time cosmic acceleration. For the sake of heuristic
 argument let us note that gravitational potential for a static
 point source in case of massive graviton with mass $m$ is given by,
 $-GM e^{-mr}/r$ with $m\sim H_0$ which reduces to Newtonian potential for
 $mr<<1$. However, at large scales such that $mr\sim 1$, adding mass to graviton  gives rise
 to
 weakening of gravity. Thus the introduction of mass
 is effectively equivalent to repulsive effect {\it a la}
 cosmological constant in the standard lore. It is broadly clear
 that cosmological constant gets linked to graviton mass which is
 altogether a novel perspective. Secondly, one might have a naive feeling that since the mass of graviton
is very small, the Pauli-Fierz theory would not disturb the
predictions of GR in the local neighborhood. The deep scrutiny of
the problem reveals that it does not hold and that the predictions
Pauli-Fierz theory in the solar system are at finite difference from
GR, hence the theory suffers from discontinuity problem dubbed vDVZ
discontinuity \cite{vanDam:1970vg,Zakharov:1970cc}.
In what follows, we shall outline the problem and
expose its underlying cause.
\subsection{vDVZ discontinuity}
\begin{figure}[h]
\includegraphics[scale=0.4]{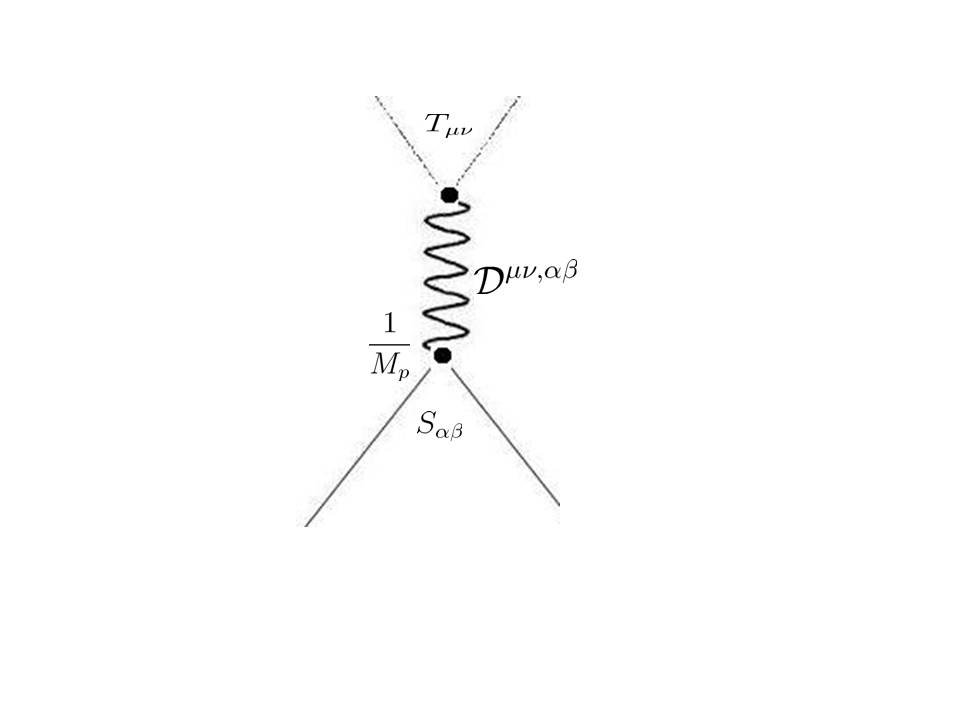}
\caption{Tree level scattering of two matter sources $T_{\mu\nu}$
and $S_{\alpha \beta}$ which couple with the universal coupling
constant, $1/M_p$  and $\mathcal{D}^{\mu\nu,\alpha\beta}$ is the
propagator of massive(massless) graviton} \label{massgr}
\end{figure}

As mentioned before, the field $h_{\mu\nu}$ universally couples to
any matter source $T_{\mu\nu}$. If we expand the Einstein-Hilbert
term in presence of a matter Lagrangian up to leading order in
$h_{\mu\nu}$, we not only reproduce $\mathcal{L}_{m=0}$ but also
obtain the coupling of the field with the source, namely, $
h_{\mu\nu}T^{\mu\nu}/M_p$. Hence the Lagrangian of the massive
spin$-$2 field interacting with matter source has the following
form \cite{Hinterbichler:2011tt},
\begin{eqnarray}
\label{pfc} \mathcal{L}=\frac{1}{2}
\partial_\lambda h_{\mu\nu}
\partial^\lambda h^{\mu\nu}+\partial_{\mu} h_{\nu\lambda}
\partial^\nu h^{\mu\lambda}-\partial_\mu h^{\mu\nu} \partial_\nu
h+\frac{1}{2}
\partial_\lambda h \partial^\lambda
h+\frac{1}{2}m^2\left(h_{\mu\nu}h^{\mu\nu}-h^2\right)+\frac{1}{M_p}h_{\mu\nu}T^{\mu\nu}
\end{eqnarray}
In order to understand the problem, we need to compute the
scattering amplitude of two matter sources for which we need the
expressions of propagators for massless and massive
gravitons(Fig.\ref{massgr}). These propagators can be written using
the free part of (\ref{pfg}), skipping details, we quote their
expressions \cite{Hinterbichler:2011tt},
\begin{eqnarray}
\label{prop1}
&&\mathcal{D}^{0}_{\alpha\beta,\rho\sigma}=-\frac{1}{k^2}\Big[\frac{1}{2}\left(\eta_{\alpha\rho}
\eta_{\beta\sigma}+\eta_{\alpha\sigma}
\eta_{\beta\rho}\right)-\frac{1}{2}\eta_{\alpha\beta}\eta_{\rho\sigma}\Big]\\
&&\mathcal{D}^{m}_{\alpha\beta,\rho\sigma}=-\frac{1}{k^2+m^2}\Big[\frac{1}{2}\left(\eta_{\alpha\rho}
\eta_{\beta\sigma}+\eta_{\alpha\sigma}
\eta_{\beta\rho}\right)-\frac{1}{3}\eta_{\alpha\beta}\eta_{\rho\sigma}\Big]
\label{prop2}
\end{eqnarray}
It should be noticed that the numerical coefficients of last terms
in (\ref{prop1}) $\&$ (\ref{prop2}) are different. The fact, that a
massive graviton has five degrees of freedom whereas massless
graviton has only two, is reflected in the expressions of their
propagators. Let us now compute the tree level amplitude of
scattering of two matter sources $S_{\alpha\beta}$ and
$T_{\alpha\beta}$ in massless and massive gravity. The corresponding
amplitudes are given by

\begin{eqnarray}
&&A^{(0)}=-\frac{8\pi G}{k^2}\left(S_{\mu\nu}T^{\mu\nu}-\frac{1}{2}ST\right)\\
&&A^{(m)}=-\frac{8\pi
G}{k^2+m^2}\left(S_{\mu\nu}T^{\mu\nu}-\frac{1}{3}ST\right)
\end{eqnarray}
In case of two static sources with masses, $M_1$ and $M_2$, we have,
\begin{eqnarray}
&&A^{(0)}=-\frac{4\pi}{k^2}GM_1 M_2\\
&&A^{(m)}=-\frac{4\pi}{k^2+m^2}\left(\frac{4}{3} G\right) M_1 M_2
\end{eqnarray}
In mass going to zero limit $m\to 0$, the amplitude $A^{(m)}$ does
not reduce to $A^{(0)}$ as opposed to our naive expectations.
Massive gravity in $m\to 0$ goes to a theory in which $G$ gets
replaced by $4 G/3$. We therefore conclude that linear massive
gravity is at finite difference from $GR$ and hence inconsistent. In
case we deform the parameters in a theory and then switch off the
deformation, logical consistency demands that the modified theory
should reduce to the original set up which does not happen in case
of PF theory.

\begin{figure}[h]
\includegraphics[scale=0.6]{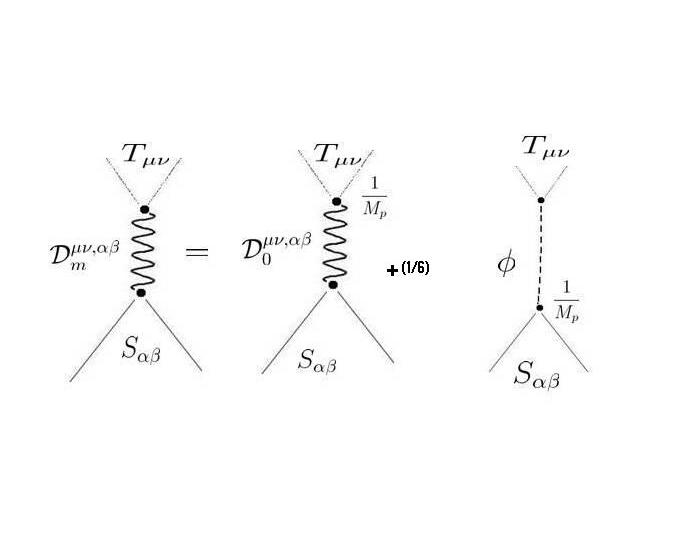}
\caption{Scattering of two matter sources $T_{\mu\nu}$ and
$S_{\alpha \beta}$ at tree level in massive gravity. The process in
the $m\to 0$ limit is represented by the exchange of massless
graviton as usual(first diagram on right) plus an extra interaction
mediated by the longitudinal mode $\phi$ which couples with the
matter sources with the universal coupling of GR.} \label{vdvz2}
\end{figure}

Before addressing the problem, we have to clearly understand the
underlying reason for vDVZ discontinuity. We shall present heuristic
arguments without going into detailed exposition of the problem.
First of all, we note that the procedure of taking limit should be
legitimate, it should preserve the degrees of freedom. The correct
frame work of carrying out such a program is provided by Stukelberg
formalism \cite{Stueckelberg:1957zz,Ruegg:2003ps} which reinstate
the gauge invariance broken by Pauli-Fierz mass term. After taking
then the $m\to 0 $ limit, we have to worry about the three extra
degrees of freedom. In case of massive vector field, the extra
(longitudinal) degree of freedom gets decoupled from the system
thereby no discontinuity problem. Let us recall that in case of the
Yang-Mills, say SU(2), theory, if one of vector bosons happens to be
in the longitudinal state, it can be decoupled from the system
whereas the other two can not be; in this case one requires Higgs
field to address the problem. It is therefore quite possible that
the extra degrees of freedom in case of gravity might not  decouple
from the source. Let us write the following decomposition for
$h_{\mu\nu}$,
\begin{equation}
h_{\mu \nu}=h^t_{\mu\nu}+\partial_\mu A_\nu+\partial_\nu
A_\mu+\partial_\mu\partial \nu \phi
\end{equation}
Such a decomposition can be understood either from group
representation or at the level of Lagrangian
formalism\cite{Hinterbichler:2011tt}. In $m\to 0$ limit,
$h^t_{\mu\nu}$, $A_\mu$ represent two transverse degrees of massless
graviton, two degrees of freedom of massless vector field whereas
$\phi$ is the longitudinal component of $h_{\mu\nu}$. Let us argue
that $A_\mu$ will not couple with the given conserved source
$T_{\mu\nu}$.  Its coupling could be of the form, $(\partial_\mu
A_\nu+\partial_\nu A_\mu)T^{\mu\nu}$; by integration of parts, we
can through the derivative on $T_{\mu\nu}$ to discard this
possibility. As for the longitudinal component, the only possibility
is that it couples with the trace of $T_{\mu\nu}$ as $\phi T$. The
detailed investigations reveal that indeed this is the case and the
coupling constant is same as in case of the massless
graviton\cite{Hinterbichler:2011tt}. In massive gravity, there is an
extra contribution to the scattering amplitude due to the exchange
of scalar degree which is of the same order as the amplitude in
Einstein gravity. This could also be noticed by rewriting the
propagator of massive graviton in the following form,
\begin{equation}
\mathcal{D}^{m}_{\alpha\beta,\rho\sigma}=-\frac{1}{k^2+m^2}\Big[\frac{1}{2}\left(\eta_{\alpha\rho}
\eta_{\beta\sigma}+\eta_{\alpha\sigma}
\eta_{\beta\rho}\right)-\frac{1}{2}\eta_{\alpha\beta}\eta_{\rho\sigma}\Big]+\frac{1}{6}
\frac{\eta_{\alpha\beta}\eta_{\rho\sigma}}{k^2+m^2}
\end{equation}
where the first term represents transverse part of massive spin-2
field propagator whereas the second part is nothing but propagator
of massive scalar field. It is therefore clear that the theory under
consideration can not reduce to GR, see Fig.\ref{vdvz2}).\\

We exposed the underlying reason of the discontinuity which is
generic to linear massive gravity {\it a la} Pauli-Fierz . How do we
cure this problem. The irony is that again we deal with an extra
scalar field similar to the chameleon theory. In that case we
implemented chameleon screening which is not viable in this case as
the scalar degree of freedom is massless. It was pointed out by
Vainshtein in 1972 that the linear approximation breaks down in the
neighborhood of a massive body below certain radius $r_V$ and that
the non linear effects screen out any modification to gravity below
$r_V$ leaving GR intact there. It tempting to think that the
longitudinal degree of freedom could be galileon though this aspect
of Vinshtein screening became known very recently. Actually, this
mechanism is in built in DGP \cite{Dvali:2000hr} where lowest order
Galileon term occurs in the so called {\it decoupling limit}. The
connection of Galileon to screening was the central point in the
formulation of dRGT. Before we discuss this development, let us show
that PF theory will have ghost if we try to extend it to non linear
background or we break the Pauli-Fierz tuning. In both the cases, we
end up with equations of motion of order higher than second which
inevitably leads to Ostrogradki instability or ghosts.
It is not by chance that first evolution equation $-$ the Newton's
second law is a second order equation. We should wonder why
dynamical equations that we come across are of second order. The
answer to this profound question was provided by Ostrogradski. If
the higher order time derivative Lagrangian is non-degenerate, there
is at least one linear instability in the Hamiltonian of this system
which means that Hamiltonian is unbounded from below. In general, if
the Lagrangian is not invertible, there are constraints in the
system and Ostrogradski theorem does not hold; such a system might
be stable. The Ostrogradski  Lagrangian essentially leads to
equations of motion of higher order than second. While quantizing a
system whose Hamiltonian is unbounded from below, one encounters
negative norm states dubbed {\it ghosts} \cite{Ostrogradski}.
\subsection{Ostrogradski $(ghosts)$ instability }
In order to see how Ostrogradski instability occurs, let us for
simplicity consider a Lagrangian, $\mathcal{L}(q,\dot{q})$ with the
standard equation of motion(see Ref.\cite{Woodard}, an excellent
review on this theme),
\begin{equation}
\frac{\partial \mathcal{L}}{\partial q}-\frac{d}{dt}\frac{\partial
\mathcal{L}}{\partial \dot{q}}=0
\end{equation}
The Lagrangian is non degenerate if $\frac{\partial^2
\mathcal{L}}{\partial \dot{q}^2}\neq 0$ which simply means that
$\frac{\partial \mathcal{L}}{\partial \dot{q}}$ depends upon
$\dot{q}$. In this case, Lagrangian equation can be cast in the form
of Newton's second law ,
\begin{equation}
\ddot{q}=f(q,\dot{q})
\end{equation}
 whose unique solution $q(t)$ requires the knowledge of two initial
 conditions on $q(t)$ and $\dot{q}(t)$. We can then transform from
 configuration space $(q,\dot{q})$ to phase space $(q,p)$
by defining the canonical momentum $p$,
\begin{equation}
p=\frac{ \partial \mathcal{L} }{\partial \dot{q}}
\end{equation}
which thank to the non degeneracy of the Lagrangian allows us to
express, $\dot{q}$ in terms of $q$ and $p$.
  One then sets up
 the Hamiltonian
\begin{equation}
H(q,p)=p\dot{q}-\mathcal{L}(q,\dot{q})\to dH=\dot{q}dp-\dot{p}dq
\end{equation}
 from which reads out the Hamiltonian equations,
\begin{equation}
\dot{q}=\frac{\partial H}{\partial p},~~~\dot{p}=-\frac{\partial
H}{\partial q}
\end{equation}
that are  equivalent to Lagrangian
 equation.
Things would qualitatively
 change in case the Lagrangian depends on time derivatives higher
 than one. Indeed, let us consider the Lagrangian,
 $\mathcal{L}(q,\dot{q},\ddot{q})$ for which the  equation of motion
 has the following form,
 \begin{equation}
 \label{lhd}
\frac{\partial \mathcal{L}}{\partial q}-\frac{d}{dt}\frac{\partial
\mathcal{L}}{\partial \dot{q}}+\frac{d^2}{dt^2}\frac{
\partial\mathcal{L}}{ \partial \ddot{q}}=0
 \end{equation}
 In this case, the non degeneracy of the Lagrangian would imply that
 $\frac{\partial \mathcal{L}}{\partial \ddot{q}}$ depends upon
 $\ddot{q}$. Lagrangian equation (\ref{lhd}) then gives rise to
 following fourth order differential equation,
 \begin{equation}
 \label{4eq}
 \ddddot{q}(t)=f(q,\dot{q},\ddot{q},\dddot{q})
 \end{equation}
 Uniqueness of the solution $q(t)$  of  (\ref{4eq}) would require
 the extra information on the initial values of $\ddot{q}$ and
 $\dddot{q}$ in addition to $(q_0,\dot{q}_0)$. The  extra  information
 brings in instability in the system or ghost.
Indeed, analogous to the standard case, we  have four canonical
variables in this case. Following Ostrogradski, we choose them as,
 \begin{eqnarray}
&&q_1=q,~~~p_1=\frac{\partial \mathcal{L} }{\partial
\dot{q_1}}-\frac{d}{dt}\frac{\partial \mathcal{L} }{\partial
\ddot{q}}\\
&& q_2=\dot{q},~~~p_2=\frac{\partial \mathcal{L} }{\partial
\ddot{q}}
 \end{eqnarray}
 Non degeneracy of the Lagrangian means that we can express $\partial
 \mathcal{L}/\partial \ddot{q}$ through $q_1,q_2$ and $p_2$. We can
 then set up the Ostrogradski Hamiltonian,
\begin{eqnarray}
\label{osh}
 H= p_1\dot{q}_1+p_2\dot{q}_2-\mathcal{L}(q_1,q_2,p_2)=p_1q_2+p_2\dot{q}_2-\mathcal{L}(q_1,q_2,p_2)
\end{eqnarray}
The Hamiltonian equations for (\ref{osh}) analogous to the standard
case have the similar form,
\begin{equation}
\dot{q}_i=\frac{\partial H}{\partial p_i};~~\dot{p}=-\frac{\partial
H}{\partial q_i},~~i=1,2
\end{equation}
 and it is not difficult to check that they are equivalent to (\ref{lhd}) and also reproduce phase space transformation.
 The  Hamiltonian (\ref{osh}) acquires a strange
piece with respect to the canonical momentum $p_1$ which primarily
appears due to the higher derivative term in the Lagrangian. As the
Hamiltonian (\ref{osh}) is linear in $p_1$, the dynamical system
under consideration is unstable in half of the phase space. Bringing
in one higher derivative brings in one bad degree of freedom. In
case, the Lagrangian contains $n$ higher derivative and satisfies
the condition of non degeneracy, the Ostrogradski Hamiltonian would
be linear in all the $n$ momenta and hence not bounded from below
along $n$ directions.
 The
Ostrogradski instability is a very generic phenomenon which can not
be cured by passing to the quantum theory. Efforts of quantizing
such a system gives rise to the negative norm states or {\it
ghosts}. By adding constraints to the system, one can not get rid of
these ghosts. One should either avoid higher order equations or
ensure that ghosts do not occur below the cut off in the effective
theory of interest hoping that UV completion would address the
problem.
\subsection{Ghosts in massive gravity}
The choice of PF mass term is very generic, as mentioned before, the
violation of Pauli-Fierz tuning leads to ghost. Indeed let us
consider the following mass term,
\begin{equation}
\mathcal{L}_m=-\frac{1}{2}m^2\left(ah_{\mu \nu}h^{\mu
\nu}-h^2\right)
\end{equation}
where a is constant. The Lagrangian $\mathcal{L}$ is invariant under
the following transformation pattern like a gauge transformation,
\begin{equation}
h_{\mu \nu}\to h_{\mu\nu}+\partial_\mu A_\nu+\partial_\nu
A_\mu+\partial_\mu\partial \nu \phi
\end{equation}
but the massive term breaks this invariance and we get an additional
term for $\phi$ field(vector field is not of interest here) that
includes higher derivative term $(\Box \phi)^2$. As a result the
$\phi$ Lagrangian we deal with in this case is of the following type
\begin{equation}
\label{pg}
 \frac{m^2}{2 d^2}(\left(\Box \phi\right)^2-\frac{1}{2}(\partial _\mu
 \phi)^2,~d^2=\frac{1}{2(a-1)}
\end{equation}
The first term in this expression is dangerous, this is the higher
order derivative term which leads to ghost in accordance with the
Ostrogradski theorem. Lets us compute the mass of the ghost. One can
easily check that (\ref{pg})  is equivalent to the following
Lagrangian\cite{Creminelli:2005qk},
\begin{equation}
\label{pg1} \mathcal{L}_g=-\frac{1}{2}(\partial_\mu
\phi)^2-m^2\partial_\mu \chi\partial^\mu\phi-\frac{1}{2}m^2d^2\chi^2
\end{equation}
where $\chi$ is an auxiliary field. Varying (\ref{pg1}) with respect
to $\chi$, one finds, $\chi=\frac{1}{d^2}\Box\phi$ and invoking it
back into (\ref{pg1}), one reinstates the original expression
(\ref{pg}). Next, changing the variable $\phi\to \phi'-m^2\chi$, one
can diagonalize (\ref{pg1}),
\begin{equation}
\label{pg2}
\mathcal{L}_g=-\frac{1}{2}(\partial_\mu\phi')^2+(\partial_\mu
\chi)^2-\frac{1}{2}m^2d^2\chi^2
\end{equation}
It is now clear that (\ref{pg}) describes two degrees of freedom,
one of which ($\chi$) is ghost and its mass is given by
\begin{equation}
m^2_{ghost}= \frac{m^2}{2(a-1)}
\end{equation}
which is infinite in case $a=1$, thereby ghost does not propagate in
Pauli-Fierz theory, it is led to sleep in its grave. Hence PF tuning
is generic for the theory to be a consistent field theory.\\
Let us now check that ghost dubbed Boulware-Deser ghost will wake up
if we try to naively extend the theory to non linear background
\cite{Boulware:1974sr}. In this case the Einstein equations take the
following form\cite{Creminelli:2005qk},

\begin{equation}
\label{eeqnl}
G_{\mu\nu}-\frac{1}{2}m^2\Big[\left(h_{\mu\nu}-h\eta_{\mu\nu})+\mathcal{O}(h^2_{\mu\nu}\right)\Big]=0
\end{equation}
where we have allowed non linearity in the mass term also. Gauge
invariance then ensures the Bianchi identity
\begin{equation}
\label{bi}
 \nabla^\mu(h_{\mu\nu}-h\eta_{\mu\nu})+\mathcal{O}(h^2_{\mu\nu})=0
\end{equation}
which forces the trace of $G_{\mu\nu}$
\begin{equation}
\label{tg}
G^\mu_\mu(L)=2\partial^\mu\partial^\nu(h_{\mu\nu}-h\eta_{\mu\nu})
\end{equation}
to vanish at the linear order. This in turn follows from
(\ref{eeqnl}) at the linear level that $h=0$. It is therefore clear
that the constraint that trace of $h_{\mu\nu}$ vanishes is specific
to linear theory. If we allow non linearity of lowest order, instead
of constraint we get an equation,
\begin{equation}
\mathcal{O}(\partial^2h^2_{\mu\nu})-\frac{1}{2}m^2\left(-3h+\mathcal{O}(h^2_{\mu\nu})\right)=0
\end{equation}
and as a consequence,  Boulware-Deser ghost becomes alive and begins
to propagate. Now we get into a dilemma, { \it linear theory has no
ghost but plagued with vDVZ discontinuity which can be resolved by
extending the theory to non linear background but the latter makes
the ghost alive}. At the onset it sounds like a no go theorem. This
is the reason why massive gravity did not progress for a long time.
Very recently, a non linear generalization  of PF theory was
proposed.
\section{$dRGT$ at a glance}
The above discussion shows that that the extension of PF theory to
non linear background leads to ghost. Question then arises, can we
generalize PF mass term  higher order than second such that ghost
does not occur. The answer is yes$-$ a very specific structure can
do that and the framework is known as $dRGT$
\cite{deRham:2010ik,deRham:2010kj}. Since the PF mass term breaks
gauge invariance, the first step is to  reinstate the general
covariance which is done by using the Stuckelberg formalism
\cite{Stueckelberg:1957zz,Ruegg:2003ps}. One needs to replace
$h_{\mu\nu}$ by a general covariant tensor $H_{\mu\nu}$; we need
four scalar fields, $\phi_a, a=1,4$,
\begin{equation}
 \label{refmet} H_{\mu\nu}=g_{\mu\nu}-\partial_\mu \phi^a
\partial_\nu \phi^b\eta_{ab}
\end{equation}
An important comment about reinstating the general covariance in
(\ref{refmet}) is in order. We could replace flat metric $\eta_{ab}$
in (\ref{refmet}) by any other metric to serve the purpose. But
changing metric would change the underlying physics. There is no
fundamental principle that can allow us to make a particular choice
except considerations based upon simplicity or phenomenology. One
way out is to turn the reference metric into dynamical one and opt
for bi-gravity theories. Let us also note that
 $\phi^a$ is scalar under diffeomorphizm but transforms
as a vector under Lorentz transformation. The PF mass term then
becomes,
\begin{equation}
\label{pfg}
 -\frac{m^2M^2_P}{8}g^{\mu
\nu}g^{\rho\sigma}\left(H_{\mu\rho}H_{\nu\sigma}-H_{\mu}H_{\rho\sigma}\right)
\end{equation}
which is the right object to cast in the non linear background,
\begin{equation}
\label{pfnl}
 \mathcal{L}=\frac{M^2_p}{2}R-\frac{m^2M^2_P}{8}g^{\mu
\nu}g^{\rho\sigma}\left(H_{\mu\rho}H_{\nu\sigma}-H_{\mu}H_{\rho\sigma}\right)+\mathcal{L}_m
\end{equation}
 where we have added the matter Lagrangian. Let
us notice that in the unitary gauge, $\phi^a=x^a$, (\ref{pfg})
reduces to PF mass term. One can go beyond unitary gauge and write
helicity decomposition for $\phi^a$ using canonical fields,
\begin{equation}
\label{helicity}
 \phi^a=x^a+{A^a}+\eta^{ab}{\partial _b\phi}
\end{equation}
As mentioned before, we can ignore the vector field and focus on
helicity zero component,
\begin{equation}
H_{\mu\nu}={h_{\mu\nu}}+2\partial_\mu\partial_\nu
\phi-\eta^{\alpha\beta}\partial_\mu\partial_\nu
\phi\partial_\alpha\partial_\beta \phi
\end{equation}
which is justified  in a limit known as decoupling limit.
\subsection{Decoupling limit}
The decoupling limit is some sort of high energy limit. This limit
is very helpful in counting the degrees of freedom in massive
gravity and provides with a valid framework to discuss the mass
screening in the local environment.

In this  limit one is dealing with energies much higher than the
mass of the graviton
\cite{ArkaniHamed:2002sp,Luty:2003vm,deRham:2010ik,deRham:2010kj},
\begin{equation}
M_p\to \infty,~m\to 0,~ T=\infty,~
\Lambda={\rm fixed},~~\frac{T}{M_p}={\rm fixed}
\end{equation}
In the decoupling limit, the dominant $\phi$ interactions survive
and Einstein-Hilbert action linearizes in $h_{\mu\nu}$ such that
(\ref{pfnl}) reduces to the following
\cite{deRham:2010ik,deRham:2010kj,Hinterbichler:2011tt},
\begin{equation}
\label{gdc}
\mathcal{L}=\mathcal{L}_{m=0}-\frac{1}{4}F_{\mu\nu}F^{\mu\nu}-
3(\partial_\mu \phi)^2+\frac{1}{\Lambda_5^5}\left[(\Box \phi)^3-\Box
\phi\partial_\mu\partial_\nu\phi)^3\right]+\frac{1}{M_p}h_{\mu\nu}T^{\mu\nu}
\end{equation}

where $\Lambda_5=(m^4M_p)^{1/5}$ is the cut off in the theory. A
comment about the decoupling Lagrangian (\ref{gdc}) is in order. We
first carry out expansion around the unitary gauge (\ref{helicity})
and the expansion around flat space time,
$g_{\mu\nu}=\eta_{\mu\nu}+h_{\mu\nu}$.

 The Einstein-Hilbert
action expanded to quadratic order in $h_{\mu\nu}$ gives rise to
$M^2_p/4$ multiplied by the quadratic piece in $h_{\mu\nu}$. The
expansion of matter Lagrangian produces, $ h_{\mu\nu}T^{\mu\nu}/2$.
Once we opt for the canonical fields, $h_{\mu\nu}\to
2h_{\mu\nu}/M_p$, we obtain $\mathcal{L}_{m=0}$ from
Einstein-Hilbert term plus the last term in (\ref{gdc}) from
$h_{\mu\nu}T^{\mu\nu}/2$. As for the higher order terms in
$h_{\mu\nu}$, they drop out in the decoupling limit. Hence
Einstein-Hilbert action linearizes in the decoupling limit.

Next by invoking the helicity decomposition in the mass  term and
using the canonical normalization for $A_\mu\to 2 A/mM_p $ and
$\phi\to 2\phi/m^2M_p$, we obtain other terms in (\ref{gdc}) in the
decoupling limit\footnote{ Actually, $\phi$ is mixed with
$h_{\mu\nu}$ and
 one needs to invoke conformal transformation to
diagnolize the degrees of freedom. $\mathcal{L}_{m=0}$ is not
invariant under conformal transformations, it gives rise to
$(\partial_\mu \phi)^2$ term in (\ref{gdc})}. Let us first note that
$\phi$ coupling with matter source survives the decoupling limit
whereas the vector field coupling does not. The non linear
derivative $\phi$ self coupling is controlled by the cut off
$\Lambda_5$ which is precisely the scale at which non linearities
become important. The non linear coupling is responsible for
restoring GR below Vainshtein radius defined by the cut off.
 Unfortunately , the higher order derivative terms in (\ref{gdc})  are
 dangerous; they do not belong to the class of galileons and give
rise to Ostrogradski ghost. Thus, $\phi$ acquires an additional
degree of freedom, a ghost which precisely cancels the contribution
of longitudinal degree of freedom and restores $GR$ within
Vainshtein radius in the non linear
background\cite{Hinterbichler:2011tt}. We thus solve $vDVZ$
discontinuity but a ghost gets introduced in the process which is
unacceptable. It would have been really remarkable, had the higher
order derivative $\phi$ Lagrangian in the decoupling limit were a
galileon Lagrangian!. The question then arises, can we include
higher order terms in the Lagrangian (\ref{pfnl}) such that ghosts
do not occur. We consider the following generalization
\cite{deRham:2010kj},
\begin{equation}
\mathcal{L}=\frac{M_p^2}{2}R-\frac{1}{8} m^2M^2_p
\mathcal{U}(H_{\mu\nu}, g_{\mu\nu})
\end{equation}
The expansion of $\mathcal{U}$ in $H_{\mu\nu}$ in its lowest order
will produce (\ref{pfnl}). However, in the nth order it will give
rise to terms like $(\partial^2\phi)^{2n}$ and would lead to ghosts
in general in view of Ostrogradski theorem. Hence, $\mathcal{U}$
should be chosen in a very specific and clever manner such that in
the decoupling limit, the non linear Lagrangian either reduces to
total derivatives or to the galileons. It is easier first to check
it in the
decoupling limit and then generalize the result beyond this limit. \\

The goal is achieved if the  structure of $\mathcal{U}$  is such
that when expanded  in $h_{\mu\nu}$,
\begin{equation}
\label{udc} \mathcal{U}_{M_p\to \infty}\equiv
\mathcal{U}_{h_{\mu\nu}\to 0}=\mathcal{U}_0+F_G(\phi) h_{\mu\nu}+...
\end{equation}
the zeroth order term, $\mathcal{U}_0$ is a total derivative and
does not reflect on the equations of motion. The first order
correction is important in the expansion (\ref{udc}), the only
option for it not to be dangerous is that $F_G(\phi)h_{\mu\nu}$
should be represented by a galileon field alone once the Lagrangian
is diagonalized thereby ghost free despite being higher derivative.
As for the higher order terms in the expansion (\ref{udc}), the same
should keep repeating. Actually, $dRGT$ operates with a specially
chosen form of $\mathcal{U}$ which satisfies this criteria. It
becomes cumbersome to tackle the higher order terms in the expansion
of $\mathcal{U}$; one can then work in the unitary gauge to confirm
that the sixth degree of freedom is indeed absent in
$dRGT$.\\
Let us now specify the form of $\mathcal{L}$,
\begin{align}
\label{actionmass} \mathcal{S}_{\rm{mass}}&=\frac{m^2
M_{Pl}^2}{8}\int {\rm d}^4 x\sqrt{-g}~ \Bigl[\mathcal{U}_2+\alpha_3
\mathcal{U}_3+\alpha_4 \mathcal{U}_4\Bigr]
\end{align}
In action (\ref{actionmass}), $\alpha_3$, $\alpha_4$ are two
arbitrary parameters and $U_i$ are specific polynomials of the
matrix
\begin{equation}
\label{Kdef} \mathcal{K}^\mu_\nu=\delta^\mu_\nu-\sqrt{g^{\mu\alpha}
\partial_\alpha\phi^a\partial_\nu\phi^b\eta_{ab}},
\end{equation}
given by,
\begin{align}
\label{us}
\mathcal{U}_2 &= 4([\mathcal{K}]^2-[\mathcal{K}^2])\\
\mathcal{U}_3 &=[\mathcal{K}]^3-3 [\mathcal{K}] [\mathcal{K}^2]+2 [\mathcal{K}^3]\\
\mathcal{U}_4 &= [\mathcal{K}]^4-6 [\mathcal{K}]^2 [\mathcal{K}^2]+3
[\mathcal{K}^2]^2+8 [\mathcal{K}] [\mathcal{K}^3]-6 [\mathcal{K}^4].
\end{align}
In  (\ref{Kdef}), $\eta_{ab}$ (Minkowski metric) is a reference
metric and  $\phi^a(x)$ are the St\"{u}ckelberg scalars introduced
to restore general covariance \cite{ArkaniHamed:2002sp}. Let us
comment on the choice of the action. We first note that  in the
 decoupling limit, the specially constructed matrix,
$\mathcal{K}|_{h_{\mu\nu}\to 0}\to \partial_\mu\partial_\nu\phi$ and
by virtue of special construction (\ref{us}), all the $\mathcal{U}$
turn into total derivatives in this limit. Actually,
$\mathcal{U}_i,i=1,3$ are only non trivial total derivatives in four
dimensions one can (uniquely) construct from
$\partial_\mu\partial_\nu \phi$ The fact that $\mathcal{U}_2$
reduces to total derivative in the zero order of the decoupling
limit speaks of the success of PF theory.  Next, one can show that
galileons occur in the first order correction when we expand the
mass term in $h_{\mu\nu}$ and diagonalize the Lagrangian using the
conformal transformation on $h_{\mu\nu}$. This ensures that local
physics is taken care of by the Vainshtein effect and no ghost
occurs in the decoupling limit. It can be demonstrated that this
result remains valid beyond the decoupling limit.

\subsection{FRW cosmology: Difficulties of $dRGT$}
We consider a flat  Friedmann-Robertson-Walker (FRW) metric of the
form\cite{massc}
\begin{align}
{\rm d}s^2=g_{\mu\nu}d x^\mu d x^ \nu=-N(t)^2dt^2+a^2(t)\delta_{ij}
dx^i dx^j
\end{align}
while for the  St\"{u}ckelberg scalars we consider the ansatz
\begin{align}
\phi^0=f(t), ~~~ \phi^i=x^i.
\end{align}
In this case the Einstein-Hilbert action and the action \ref{actionmass} become:
\begin{align}
\mathcal{S}_{\rm{EH}}&=-3M_{Pl}^2\int {\rm d}t \Bigl[\frac{a\dot
a^2}{N}\Bigr]\\
\label{gravitonpart}
\mathcal{S}_{\rm{mass}}&=3m^2 M_{Pl}^2\int {\rm d}t a^3\Bigl[N G_1(\xi)-\dot f a G_2(\xi)\Bigr]\\
\end{align}
where we have defined
\begin{align}
\label{G1def} G_1(\xi) &= (1-\xi)\Bigl[2-\xi+\frac{\alpha_3}{4}
(1-\xi)(4-\xi)+\alpha_4
(1-\xi)^2\Bigr]\\
\label{G2def} G_2(\xi) &= \xi (1-\xi)\Bigl[1+\frac{3}{4}\alpha_3
(1-\xi)+\alpha_4 (1-\xi)^2\Bigr] ; \xi = \frac{1}{a}
\end{align}

Variation with respect to $N$, setting $N=1$ at the end, leads to
the first Friedmann equation
\begin{align}
\label{eq:Fried} 3M_{Pl}^2H^2={\rho_m+\rho_r-3 m^2 M_{p}^2G_1}
\end{align}

Variation with respect to $f$  gives the constraint equation
\begin{align}
\label{eq:Constraint} \frac{d}{dt}\left(m^2a^4G_2(a)\right)=0~\to
G_2(a)=\frac{C}{a^4}
\end{align}

where $C$ is a constant of integration. The constraint equation is
problematic as it implies that $a=const$. We can invoke spatially
non flat geometry and obtain a viable background dynamics. However,
the latter turn out to be unstable under perturbations\cite{ma}.
Even if these branches were stable, the absence of spatially flat
geometry would mean point towards some underlying problem of $dRGT$.
There are clearly three ways to handle the problem: (1) We may
adhere to the point of view that the mass of graviton is strictly
zero and abandon the efforts to look for consistent theory of
massive gravity. (2) Perhaps the easiest way out is to modify $dRGT$
at the level of cosmology, say by making the mass of graviton a
field variable by replacing $m\to V(\psi)$ or by introducing a
dilaton field which by itself can not lead to a fundamental
idea\cite{dilaton}. This is similar to curing a wound from out side
without providing an internal therapy. The bi-metric theories and
the models of massive gravity based upon Lorentz violations could be
a serious option in this category\cite{defa,hr,rubakov}. (3) If we
adopt a conservative and pragmatic view , we might claim that $dRGT$
is a correct theory. It predicts a generic anisotropy in the
universe and points towards the violation of cosmological principle.
The recent investigations on optical polarizations, CMB quadrupole
and octopole and the study of radio sources point towards a large
scale anisotropy with the preferred axis(see, Ref.\cite{jain} and
references therein). We may therefore abandon the FRW cosmology and
opt for an anisotropic background.\footnote{We thank S. Mukohyama
for highlighting this point.The prejudice against this view is
clearly associated with the fact that most of the successes of the
standard model of universe are related to the homogenous and
isotropic geometry and perturbations around it. The paradigm shift
obviously causes a resentment.} (4) The most challenging way out is
to modify $dRGT$ at the fundamental level. Let us note that
bi-gravity theories sound promising with healthy FRW cosmology  at
late times. Unfortunately, the theory runs into difficulties in the
early universe\cite{maeda}.
\section{summary and outlook}
In this brief review we presented a broad account of standard lore
of cosmic acceleration {\it a la} dark energy and large scale
modification of gravity. Given the observational constraints and
difficulties associated with model building of dynamical dark
energy, it would be fair to say that cosmological constant emerges
in a stronger position. There are three distinguished features which
make it a celebrity. First, it is the integral part of Einstein
gravity and requires no adhoc assumption for its introduction. In
fact, it makes classical Einstein gravity complete in four
dimensions. Secondly, it provides with the simplest possibility to
describe late time cosmic acceleration. Last but not least, it is
consistent with all the observations and performs better than models
of dynamical dark energy. Given the present data which is quite
accurate at the background level, we can not distinguish
cosmological constant from quintessence or large scale modification
 ; the needle of hope points towards the
 cosmological constant as the source of cosmic acceleration.
 At present, the cosmology community tacitely agrees that at the background level
 there is nothing but cosmological constant. However,one should admit that there are
 difficult theoretical issues associated
with cosmological constant. Its incredibly small value and the
absence of a generic symmetry at the associated energy scales to
protect it from quantum corrections make the problem most
challenging in theoretical physics.\\
With a hope to alleviate the cosmological constant problem, a
variety of scalar field models were  introduced. We have here
presented basic features of cosmological dynamics of scalar fields.
In our opinion, scalar field models can not address the said
problem. A fundamental scalar field is plagued with naturalness
problem thereby one problem translates into another one of similar
nature\cite{smyr}.
\\
There is a school of thought in cosmology which preaches   the
necessity of paradigm shift, namely, that large scale  modification
of gravity might account for late time cosmic acceleration. As we
pointed out earlier, the generic modifications amount to extra
degrees of freedom expected to complement Einstein gravity at large
scales. The tough challenges of these scenarios are related to local
physics constraints. In case the extra degrees are massive, the
required accuracy of their screening {\it a la chameleon mechanism}
in the local neighborhood leaves no scope for large scale
modification to account for late time
cosmic acceleration. \\
As for the massless degrees, they should be represented by galileon
field which can implement Vainshtein screening. Galileon field
appears in the decoupling limit in massive gravity. We have briefly
described the connection of galileon field  to their higher
dimensional descendent, the Lovelock gravity which leaves no
surprise for them to be ghost free. However, the fact that they can
protect local physics in the decoupling limit is a big bonus for
galileons. There are, however, issues here which need attention.
Galileons are legitimate representatives of a profound structure in
higher dimensions$-$ the Lovelock theory. The linkage of these two
systems may be established through dimensional reduction $-$ a well
defined procedure to establish contact with four dimension we live
in. It is really surprising where galileon field inherit
superluminality from. This feature can not come  from Lovelock
structure, may be it is induced from reduction process! One should
also ponder upon the connection (if any) of generalized galileon
dubbed Hordenski system to higher dimensions.\\`
 Galileon field
serves as a fundamental building block for  non linear ghost free
massive gravity. Interestingly, kicking out ghost from the theory
brings in superluminality, an inherent feature of galileons. Even if
we close our eyes on causality issues, the ghost free massive
gravity {\it a la} $dRGT$ miserably fails in cosmology it was meant
for. In our opinion, adding yet new degree to the set up, such as a
dilaton, at cosmology level defeats the original motivation of the
theory. In our description of massive gravity, we avoided technical
issues and often resorted to heuristical arguments based upon
physical perceptions. And this is consistent with the motivation of
the review to convey the basic ideas of the theme under
consideration to
a wider audience.\\
There is a beautiful field theoretic framework in the background of
the non linear ghost free massive gravity and we believe that such a
beauty can not go for waist. We believe that some fundamental idea
would resolve the underlying difficulty, may be, something similar
to the Higgs mechanism that salvaged the standard
model of particle physics.\\
On the  observational front, we  expect to distinguish between
Einstein gravity(with cosmological constant) and modified schemes
(f(R), DGP etc) in  future surveys. On theoretical grounds, the
former emerges cleaner than any large scale modification. As for the
modified theories, despite inherent difficulties, non linear massive
gravity deserves attention due to its generic features. It links
cosmological constant to the mass of a fundamental particle,  the
graviton and provides with some mechanism of degravitaion. At
present, we do not know a consistent model of large scale
modification of gravity. In such situation one might opt for an
effective description containing a single scalar field of most
general
nature non-minimally coupled to matter. We believe that future surveys of large scale structure would reveal if there
is physics beyond $\Lambda$CDM.\\
Clearly, the phenomenon of late time cosmic acceleration is far from
being understood. This is certainly the puzzle of the millennium and
it is therefore not surprising that there is no easy solution to
this problem. Observational missions are in full swing in cosmology
at present and there is no doubt that interesting times are ahead
for theoreticians as well as for observers.
\section{Acknowledgements}
We are indebted to W. Hossain, R. Gannouji , S. Mukohyama and Safia
Ahmad for taking pain in reading through the manuscript and pointing
out corrections and making suggestions for improvement. We thank  N.
Dadhich,  S. Jhingan, J. V. Narlikar, S. Odintsov, T. Padmanabhan,
V. Sahni , A. A. Sen, S. Tsujikawa and Yi Wang for comments and
fruitful discussions. MS thanks R. Kaul for fruitful discussions on
the stability of fundamental scalar under radiative corrections. He
thanks R.U. Khan for his help in drawing figures. He also thanks
Eurasian International Center for Theoretical Physics(Eurasian
National University, Astana) for hospitality where the work on the
review was completed.

\end{document}